\documentclass[twocolumn,5p]{elsarticle}
\usepackage{amsmath, amssymb}
\usepackage{graphicx,bm,hhline}
\usepackage{calrsfs}
\DeclareMathAlphabet{\pazocal}{OMS}{zplm}{m}{n}

\newcommand{\br}{{\bm r}}
\newcommand{\brp}{{\br}^\prime}

\newcommand{\hf}{\frac{1}{2}}

\newcommand\sss{\scriptscriptstyle}

\newcommand{\wig}[1]{\mathrel{\hbox{\hbox to 0pt{\lower.6ex\hbox{$\sim$}\hss}\raise.4ex\hbox{$#1$}}}}

\journal{Computer Physics Communications}

\begin{document}

\begin{frontmatter}
\title{Wide Ranging Equation of State with \texttt{Tartarus}: a Hybrid Green's Function/Orbital based Average Atom Code}

\author[lanl]{C. E. Starrett}
\ead{starrett@lanl.gov}
\author[lanl,au]{N. M. Gill}
\author[lanl]{T. Sjostrom}
\author[lanl]{C. W. Greeff}

\address[lanl]{Los Alamos National Laboratory, P.O. Box 1663, Los Alamos, NM 87545, U.S.A.}
\address[au]{Auburn University Physics Department, 206 Allison Laboratory, Auburn University, Auburn, AL 36849, USA}

\begin{abstract}
Average atom models are widely used to make equation of state tables and for calculating other
properties of materials over a wide range of conditions, from zero temperature isolated atom to fully
ionized free electron gases.  The numerical challenge of making these density functional theory based models 
work for {\it any} temperature, density 
or nuclear species is formidable.  Here we present in detail a hybrid Green's function/orbital based approach
that has proved to be stable and accurate for wide ranging conditions.  Algorithmic strategies are
discussed.  In particular the decomposition of the electron density into numerically advantageous parts 
is presented and a robust and rapid self consistent field method based on a quasi-Newton algorithm
is given.  Example application to the equation of state of lutetium (Z=71) is explored in detail, including
the effect of relativity, finite temperature exchange and correlation, and a comparison to a less approximate
method.  The hybrid scheme is found to be numerically stable and accurate for lutetium over at least 6 orders of magnitude 
in density and 5 orders of magnitude in temperature.
\end{abstract}

\begin{keyword}
Average atom \sep {\texttt Tartarus}
\end{keyword}

\end{frontmatter}


\section{Introduction}
Average atom models are computationally inexpensive tools that are used
to provide rapid equation of state and other material properties
with reasonable physical fidelity.  While more accurate models
exist, average atom models are popular not only because of
their relative rapidity, but also because they are reasonably accurate 
for a wide range of conditions, ranging from isolated atom to
free electron gas, from zero temperature to thousands
of eV.

However, while average atom models can in principle be used
for any conditions, their numerical implementation is far from
trivial.  Designing a generally robust and stable algorithm that
works for any material, or conditions, is a formidable challenge.  In this
work we discuss in some detail a hybrid orbital/Green's function
implementation that we have developed in the {\texttt Tartarus} code.

This implementation is based on the exploratory ideas presented in
references \cite{starrett15, gill17}, but builds on the much
larger base of average atom literature.  The original presentation
of the physical model used in \texttt{Tartarus} was given Liberman in references \cite{liberman, liberman82}.
This particular model was then expanded on and explored in more
detail by other authors, including references \cite{blenski, johnson06, wilson06, sterne07, bar-shalom, more85, trz18, penicaud09, ovechkin16}.
However, many other average atom like models with their own
advantages and disadvantages were also developed.  Some
include treatments of band structure (missing in Liberman's model) \cite{rozsnyai, rozsnyai01}.
Others include a more realistic treatment of ionic structure \cite{perrot90, dharma82, chihara91, starrett13}.

We present a detailed description of the model and its implementation,
including a very efficient self consistent field solution method.
We discuss the advantages of the hybrid approach and the weakness of
using purely orbitals or Green's functions.  Example is made of
the equation of state of lutetium (Z=71).  We explore the effect
of a fully relativistic treatment versus non-relativistic as well
as the effect of recent finite temperature exchange and correlation
potentials versus temperature independent potentials.
Comparison is made to a less approximate model in the low temperature
region where such models are available.
Finally, unsavory features of the model like thermodynamic inconsistency
are discussed.

\section{Average Atom Model}
\subsection{Model Description}
\begin{figure}
\begin{center}
\includegraphics[scale=0.40]{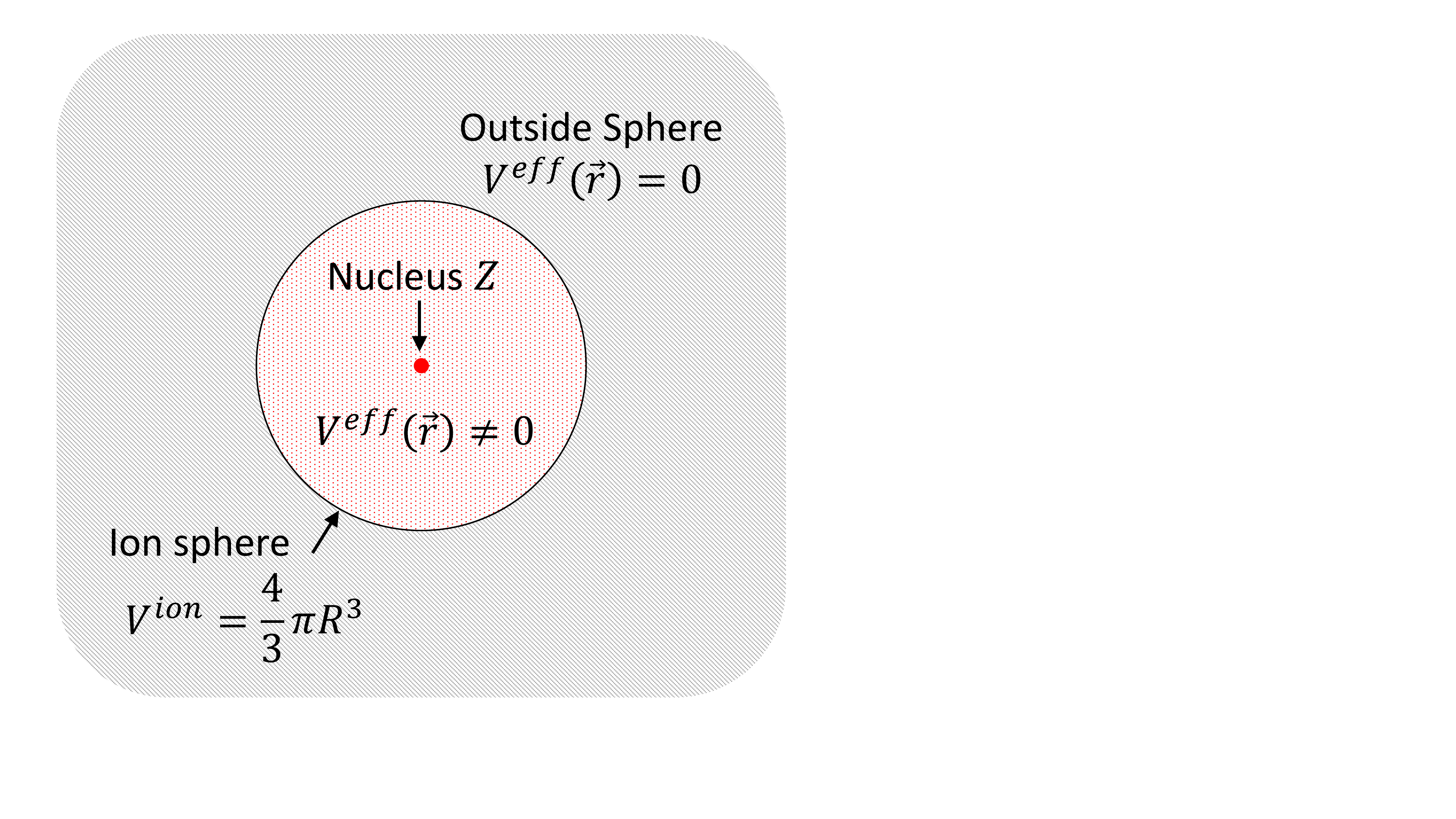}
\end{center}
\caption{(Color online) 
  Schematic diagram of average atom physical model.  Inside the ion sphere the
  electronic structure is determined with density functional theory.  The boundary
  condition is that outside the sphere the effective electron-nucleus potential
  is zero.
}
\label{aacartoon}
\end{figure}

We consider an ensemble of electrons and nuclei in local thermodynamical equilibrium.  These can form a gas, liquid, solid
or plasma.  In the average atom model we define a sphere, with a volume equal to the average volume per nucleus ($V^{ion}$), with a nucleus of
charge $Z$ placed at the center (the origin).  The sphere is required to be charge neutral and the boundary condition
at the edge of the sphere is that the effective electron-nucleus interaction potential $V^{eff}(\br) =0 $, and the electrons wavefunctions therefore match
to the known analytic solution at this boundary.  We must also set $V^{eff}(\br)=0$ outside the
sphere for reasons that will become clear later.  This situation is summarized in figure \ref{aacartoon}.

The electron density $n_e(\br)$ and $V^{eff}(\br)$ inside the sphere are determined by solving the relativistic or non-relativistic
density functional theory (DFT) \cite{mermin65, kohn65, rajagopal, macdonald} equations.  The procedure is as follows \cite{johnson06}:  starting from an initial guess at $V^{eff}(\br)$ the
Schr\"odinger or Dirac equation is solved for either the eigenfunctions $\psi_{\epsilon}(\br)$ or Green's functions $G(\br,\epsilon)$
and the electron density is constructed
\begin{eqnarray}
n_e(\br) & = & \int\limits_{-\infty}^{\infty} d\epsilon \, f(\epsilon,\mu)  \psi^\dagger_{\epsilon}(\br)  \psi_{\epsilon}(\br)  \label{ne}\\
         & = & -\frac{1}{\pi}\Im \int_{-\infty}^{\infty}d\epsilon f(\epsilon,\mu)TrG(\br,\epsilon)
\end{eqnarray}
where (non-)relativistically $G$ is ($2\times$2) $4\times4$ matrix, and $\psi$ is a ($1\times2$) $1\times4$ column vector.  The practical
formulae for evaluation of $n_e(\br)$ are given is section \ref{sec_ne}.  $f(\epsilon,\mu)$
is the Fermi-Dirac occupation factor which depends on the electron energy $\epsilon$ and chemical potential $\mu$ as
well as the plasma temperature $T$.
$\mu$ is determined by requiring the ion-sphere to be charge neutral
\begin{equation}
Z - \int_{V^{ion}} d^3r\, n_e(\br) = 0
\label{neut}
\end{equation}
With $n_e(\br)$ so determined a new $V^{eff}(\br)$ is found
\begin{equation}
V^{eff}(\br) = V^{el}(\br)+ V^{xc}(\br)
\label{veff}
\end{equation}
where the electrostatic part is 
\begin{equation}
V^{el}(\br) = -\frac{Z}{r} + \int_{V^{ion}} \,d\br^\prime\, \frac{n_e(\brp)}{\mid \br - \brp \mid} 
\label{vel}
\end{equation}
and the exchange and correlation part is
\begin{equation}
V^{xc}(\br) = \frac{\delta F^{xc}}{ \delta n_e(\br)}
\label{vxc}
\end{equation}
where $F^{xc}$ is the chosen exchange and correlation free energy.
Equations (\ref{ne}) to (\ref{vxc}) are then repeatedly solved until
self-consistent.  In section \ref{sec_scf} a rapid and robust strategy for this 
self-consistent field (SCF) problem is presented.
The system is spherically symmetric about the origin and as a result $n_e(\br) \to n_e(r)$ and
$V^{eff}(\br) \to V^{eff}(r)$.

\subsection{Poisson Equation}
Spherical symmetry simplifies the solution of the Poisson equation (equation (\ref{vel}))
\begin{equation}
V^{el}(r) = -\frac{Z}{r} + \frac{4\pi}{r}\int_0^r\,dr^\prime \,{r^\prime}^2 n_e(r^\prime) 
                             + 4\pi \int_r^R \,dr^\prime \,{r^\prime} n_e(r^\prime) 
\label{vel_ss}
\end{equation}
This result is obtained by using a Spherical Harmonic expansion of $1/|\br-\brp|$.

\subsection{Electron density \label{sec_nea}}
On applying spherical symmetry to the Dirac equation, $n_e(r)$ can be written in terms of orbitals \cite{wilson06,piron11}
\begin{eqnarray}
  n_e(r)  & = & \sum\limits_{i\in B} f(\epsilon_i,\mu) \frac{ 2|\kappa_i| } {4\pi r^2} 
            [P^{2}_{\kappa_i}(r,\epsilon_i) + Q^{2}_{\kappa_i}(r,\epsilon_i)] \nonumber\\
            && \!\!\!\!\!\!\!\!\!\!\!\!\!\!\!\!\!\!\!\!
            +  \int_{0}^{\infty}d\epsilon f(\epsilon,\mu)
            \sum_{\stackrel{\kappa = -\infty}{\kappa \ne 0}}^{+\infty}
            \frac{ 2|\kappa| } {4\pi r^2} 
            [P^{2}_{\kappa}(r,\epsilon) + Q^{2}_{\kappa}(r,\epsilon)]
\label{ne_dir_ss}
\end{eqnarray}
where the sum over $i$ runs over all bound states and
$P_\kappa(r,\epsilon)$ ($Q_\kappa(r,\epsilon)$) is the large (small) component of the radial Dirac equation.
$\epsilon$ is the energy minus the rest mass of the electron so that it is directly comparable
to the energy appearing in the Schr\"odinger equation.  
For the Schr\"odinger equation the expression for $n_e(r)$ reads
\begin{eqnarray}
  n_e(r)  & = & \sum\limits_{i\in B} f(\epsilon_i,\mu) \frac{2(2l_i+1)}{4 \pi r^2}
            [P^{2}_{l_i}(r,\epsilon_i)] \nonumber\\
   && +
  \int\limits_{0}^\infty d\epsilon\, f(\epsilon,\mu) \sum\limits_{l=0}^\infty
                  \frac{2(2l+1)}{4 \pi r^2} 
                   [P^{2}_{l}(r,\epsilon)]
\label{ne_sch_ss}
\end{eqnarray}
where $P_{l}$ is now the solution to the radial Schr\"odinger equation.
Note that the sum over $\kappa$ in equation (\ref{ne_dir_ss}) can be converted into
a sum over orbital angular momentum index $l$ with
\begin{equation}
\sum_{\stackrel{\kappa = -\infty}{\kappa \ne 0}}^{+\infty} \to 
\sum_{l=0}^{\infty}\delta_{\kappa,-l-1} + \sum_{l=1}^{\infty}\delta_{\kappa,l}
\end{equation}
where $\delta$ is the Kronecker delta.  
Using this, and setting the small components $Q_\kappa = 0$, one recovers the
non-relativistic expression (\ref{ne_sch_ss}) from the
relativsitic one (\ref{ne_dir_ss}).

In terms of the Green's function the expression for $n_e(r)$ is identical for both the relativistic and
non-relativistic cases
\begin{equation}
n_e(r) = -\frac{1}{\pi}\Im \int_{-\infty}^{\infty}d\epsilon f(\epsilon,\mu)TrG(r,\epsilon)
\label{ne_gf_aa}
\end{equation}
Relativistically the Green's function is given by
\begin{eqnarray}
\label{gf_rel_ss}
TrG(r,\epsilon) & = & -\imath p(1+\frac{\epsilon}{2mc^{2}}) 2m \nonumber\\
              &&  \!\!\!\!\!\!\!\!\!\!\!\!\!\!\!\!\!\!\!\!\!\!\!\!\!\!\!\!\!\!\!\!\!\!\!\!\!\!\!\!\times
              \sum_{\stackrel{\kappa = -\infty}{\kappa \ne 0}}^{+\infty}
              \frac{2|\kappa|}{4\pi r^2}
                 [P_{\kappa}^{R}(r,\epsilon)P_{\kappa}^{I}(r,\epsilon) + Q_{\kappa}^{R}(r,\epsilon)Q_{\kappa}^{I}(r,\epsilon)]
\end{eqnarray}
where
\begin{equation}
p = \sqrt{2m \epsilon \left( 1+\frac{\epsilon}{2mc^2}\right)},
\end{equation}
is the magnitude of momentum, 
$P^R$ ($P^I$) is the large component, regular (irregular) solution to the radial Dirac equation, and $Q^R$ ($Q^I$) the
corresponding small components (see section \ref{sec_bc}).

Non-relativistically, the trace of the Green's function becomes
\begin{equation}
\label{gf_nonrel_ss}
TrG(r,\epsilon) = -\imath p 2m \sum_{l=0}^{\infty} \frac{2(2l+1)}{4\pi r^2}[P_{l}^{R}(r,\epsilon)P_{l}^{I}(r,\epsilon)]
\end{equation}
where $P^R$ ($P^I$) is the regular (irregular) solution to the radial Schr\"odinger equation, and
\begin{equation}
p = \sqrt{2 m \epsilon }.
\end{equation}

\subsection{Boundary Conditions\label{sec_bc}}
The boundary conditions at the sphere are that the wavefunctions must match the solution
to the Dirac or Schr\"odinger equations with $V^{eff}(r) = 0$, where both equations reduce to the spherical
Bessel equation. Relativistically, for negative energy ( $\epsilon < 0 $ i.e. the bound states), the radial wavefunctions must match 
\begin{eqnarray}
  P_{\kappa}(R,\epsilon) & = & A_{\epsilon,\kappa} R\,\imath^l \,h_l(pR) \\
  Q_{\kappa}(R,\epsilon) & = & A_{\epsilon,\kappa} R\, \imath^{l} \,\mathrm{Sgn}(\kappa) \sqrt{\frac{\epsilon}{\epsilon+2m c^2}} h_{\bar{l}}(pR)
\end{eqnarray}
with $h_{l}$ the spherical Hankel function and $\bar{l}=l- \mathrm{Sgn(\kappa)}$, where $\mathrm{Sgn}$ returns the sign of the argument,
and $A_{\epsilon,\kappa}$ is a constant of proportionality that is determined by the normalization integral
\begin{equation}
\int_0^\infty dr [P^{2}_{\kappa_i}(r,\epsilon_i) + Q^{2}_{\kappa_i}(r,\epsilon_i)] = 1
\label{norm}
\end{equation}
It is for this normalization integral that we must assume $V^{eff}(r)  =0$ for $r \ge R$.
For positive energies
\begin{eqnarray}
\!\!\!\!\!  P_{\kappa}(R,\epsilon)\!\!\!\! & = & \!\!\!\! \sqrt{\frac{p}{\pi \epsilon}} pR \left[ \cos\delta_{\kappa} j_l(pR) + \sin\delta_{\kappa} \eta_l(pR) \right]\label{bc1_rel}\\
\!\!\!\!\!  Q_{\kappa}(R,\epsilon)\!\!\!\! & = & \!\!\!\! - \mathrm{Sgn}(\kappa) \sqrt{\frac{\epsilon}{\epsilon + 2mc^2}} \nonumber\\
                       && \!\!\!\!\!\!\!\!\!\!\!\!\!\!\!\!\!\times \sqrt{\frac{p}{\pi\epsilon}} pR \left[ \cos\delta_{\kappa} j_{\bar{l}}(pR) + \sin\delta_{\kappa} \eta_{\bar{l}}(pR) \right]
                       \label{bc2_rel}
\end{eqnarray}
where $j_l$ ($\eta_l$) is the spherical Bessel (Neumann) function.
$\delta_{\kappa}$ is the energy dependent phase shift.  The numerical $P_\kappa$ and $Q_\kappa$ have arbitary normalization.  To
recover the correct physical normalization (equations (\ref{bc1_rel}) and (\ref{bc2_rel})) they are mutiplied by a constant.  This
constant and $\delta_\kappa$ are determined by requiring the numerical the boundary conditions to be satisfied.

Non-relativistically, for negative energies, we have
\begin{eqnarray}
  P_{l}(R,\epsilon) & = & A_{\epsilon,l} R\, \imath^l \, h_l(pR) 
\end{eqnarray}
with $A_{\epsilon,l}$ determined by
\begin{equation}
\int_0^\infty dr [P^{2}_{l_i}(r,\epsilon_i)] = 1
\end{equation}
and for positive energies
\begin{eqnarray}
\!\!\!\!\!  P_{l}(R,\epsilon) & = &  \sqrt{\frac{2mp}{\pi}} R \left[ \cos\delta_{l} j_l(pR) + \sin\delta_{l} \eta_l(pR) \right]\label{bc_nonrel}
\end{eqnarray}
where $\delta_{l}$ and the normalization constant for the numerical $P_l$ are determined by requiring the numerical value of $P_l(r)$ and its first derivative with respect to $r$
to satisfy the boundary condition (\ref{bc_nonrel}) and its derivative.

Relativisitcally, for the regular solutions used to construct the Green's function (equation (\ref{gf_rel_ss})), the
boundary conditions are
\begin{eqnarray}
\!\!\!\!\!  P_{\kappa}^R(R,\epsilon) & = &  R \left[ j_l(pR) -\imath p  h_l(pR) t_l(p) \right]\\
\!\!\!\!\!  Q_{\kappa}^R(R,\epsilon) & = &  \mathrm{Sgn}(\kappa) \sqrt{\frac{\epsilon}{\epsilon + 2mc^2}}  \nonumber\\
                                     &   & \times   R \left[ j_{\bar{l}}(pR) -\imath p  h_{\bar{l}}(pR) t_l(p) \right]
\end{eqnarray}
where $t_l$ is the energy dependent t-matrix that is determined by matching the numerical solution to this
boundary condition.  It is worth noting that for real energies $\epsilon$ the phase shifts and the t-matrix
are simply related \cite{zabloudil_book}.
For the irregular solutions
\begin{eqnarray}
\!\!\!\!\!  P_{\kappa}^I(R,\epsilon) & = &  R h_l(pR) \\
\!\!\!\!\!  Q_{\kappa}^I(R,\epsilon) & = &  \mathrm{Sgn}(\kappa) \sqrt{\frac{\epsilon}{\epsilon + 2mc^2}} R h_{\bar{l}}(pR)
\end{eqnarray}

The boundary conditions for the non-relativistic case are
\begin{eqnarray}
\!\!\!\!\!  P_{l}^R(R,\epsilon) & = &  R \left[ j_l(pR) -\imath p  h_l(pR) t_l(p) \right]\\
\!\!\!\!\!  P_{l}^I(R,\epsilon) & = &  R h_l(pR) 
\end{eqnarray}

\subsection{Density of States}
Relativistically the density of states $\chi(\epsilon)$ in terms of orbitals is
\begin{eqnarray}
\!\!\!\!\!\!\!\!
  \chi(\epsilon)  & \!\!\!\!\!= \!\!\!\!\!&  \sum_{i\in B}\delta(\epsilon_i-\epsilon) \int_0^R dr\,  2|\kappa_i| 
            [P^{2}_{\kappa_i}(r,\epsilon_i) + Q^{2}_{\kappa_i}(r,\epsilon_i)] \nonumber\\
            &\!\!\!\!\!\!\!\!\!\!& 
\!\!\!\!\!\!\!\!
            +  \sum_{\kappa}  2|\kappa| \int_0^R dr\,
            [P^{2}_{\kappa}(r,\epsilon) + Q^{2}_{\kappa}(r,\epsilon)] \Theta(\epsilon)
            \label{ss_dos_dir}
\end{eqnarray}
where $\delta$ is the Dirac delta function, and $\Theta$ is the Heaviside step function.  Non-relativistically the density of states is
\begin{eqnarray}
  \chi(\epsilon)  & = &  \sum_{i\in B}\delta(\epsilon_i-\epsilon) \int_0^R dr\,  2(2l_i+1) 
            [P^{2}_{l_i}(r,\epsilon_i) ] \nonumber\\
            && 
            +  \sum_{l}  2(2l+1) \int_0^R dr\,
            [P^{2}_{l}(r,\epsilon)] \Theta(\epsilon)
\end{eqnarray}
In terms of the Green's function, the expression is identical for both the
relativistic and non-relativistic cases
\begin{equation}
\chi(\epsilon)=  -\frac{1}{\pi}\Im \int_{V^{ion}}d^3r\, TrG(r,\epsilon)
\label{dos_gf_ss}
\end{equation}

\subsection{Equation of State}
The electronic free energy $F$ and internal energy $U$ per atom are
\begin{eqnarray}
F & = & F^{el} + F^{xc} + F^{ks} \\
U & = & F^{el} + U^{xc} + U^{k} 
\end{eqnarray}
$F^{el}$ is the electrostatic contribution
\begin{equation}
F^{el}=\frac{1}{2}\int_{V^{ion}}d^3r \left[ V^{el}(r)-\frac{Z}{r}\right] n_{e}(r)
\label{fel}
\end{equation}
$F^{xc}$ ($U^{xc}$) is the exchange and correlation free (internal) energy and $F^{ks}$ is the kinetic 
and entropic term
\begin{equation}
F^{ks} = U^{k} - T S
\end{equation}
where $U^{k}$ is the electron kinetic energy contribution to the internal energy
\begin{equation}
U^{k}  =  \int_{-\infty}^{\infty}d\epsilon f(\epsilon,\mu)\chi(\epsilon)\epsilon - 
            \int_{V^{ion}}d^3r V^{eff}(r) n_{e}({r})
\end{equation}
and $S$ is the entropy
\begin{eqnarray}
S & = & - \int_{-\infty}^{\infty} d\epsilon 
\chi(\epsilon ) \nonumber\\
&&
\!\!\!\!\!\!\!\!\!\!\!\!\!\!\!\!\!\!\!\!
\!\!\!\!\!\!\!\!
\times \left[
f(\epsilon,\mu)\ln(f(\epsilon,\mu)) + (1-f(\epsilon,\mu))\ln(1-f(\epsilon,\mu)) 
\right]
\label{fs}
\end{eqnarray}
The electronic pressure $P$ calculated using the Virial theorem is
\begin{equation}
P = \frac{1}{V^{ion}} \left[ 
\frac{\mathcal{T} + F^{el}}{3} 
\right] + P^{xc}
\label{pvir}
\end{equation}
where
\begin{equation}
P_{xc}= \frac{1}{V^{ion}}  \left[ -F^{xc} + \int_{V}d^3r\, n_{e}(r) V^{xc}(r)\right]
\end{equation}
and $\mathcal{T}$ is
\begin{equation}
\mathcal{T}  = 
2 \int_{-\infty}^{\infty}d\epsilon f(\epsilon,\mu)\chi^A(\epsilon)\epsilon - 
            2 \int_{V^{ion}}d^3r V^{eff}(r) n^A_{e}({r})
\end{equation}
Here the superscript $A$ means the quantity due only to the large component.  For the
relativistic case this means setting $Q_\kappa = 0$ in the expressions for the
density (\ref{ne_dir_ss}) and density of states (\ref{ss_dos_dir}), and in the Green's function (\ref{gf_rel_ss}) 
which is then used in expressions (\ref{ne_gf_aa}) and (\ref{dos_gf_ss}).
For the non-relativistic case, there is no small component, so $n_e(r) = n_e^A(r)$
and $\chi(\epsilon) = \chi^A(\epsilon)$.

\subsection{Summary}
In this section we have given formulae that both define the model and can 
be use to evaluate it numerically.  In the following section we present practical strategies
for solution of the model over a wide range of densities, temperatures and
elements based on these expressions.

\section{Numerical Methods}
\subsection{Numerical Solution of the Schr\"odinger and Dirac Equations}
The radial Dirac equations are
\begin{eqnarray}
(V^{eff}(r)-\epsilon)P_\kappa +c(\frac{d}{dr} - \frac{\kappa}{r})Q_\kappa  & \!\!\!\!\!=\!\!\!\!\! & 0 \\
-c(\frac{d}{dr} + \frac{\kappa}{r})P_\kappa + (V^{eff}(r)-\epsilon-2mc^2)Q_\kappa &\!\!\!\!\! =\!\!\!\!\! & 0
\end{eqnarray}
and the radial Schr\"odinger equation is
\begin{equation}
\frac{d^2P_l}{dr^2} + 2\left( \epsilon -V^{eff}(r) - \frac{l(l+1)}{2r^2}\right) P_l = 0
\end{equation}
These can be solved numerically with a variety of methods.  We recommend using the Adams 
methods, as explained in detail in reference \cite{johnson_book} (also used in \cite{ondrej_bnd}).  We have used the fifth
order formula.  This is a predictor-corrector method, but solves the predictor-corrector
loop analytically.  A robust method for obtaining the necessary four point starting values for 
outward integration is also presented in \cite{johnson_book} and is straightforwardly adapted for the
inward integrations.  Inward integrations
(i.e. from $R$ to $0$) for bound states and the irregular solutions start from the boundary condition values.

\begin{figure}
\begin{center}
\includegraphics[scale=0.33]{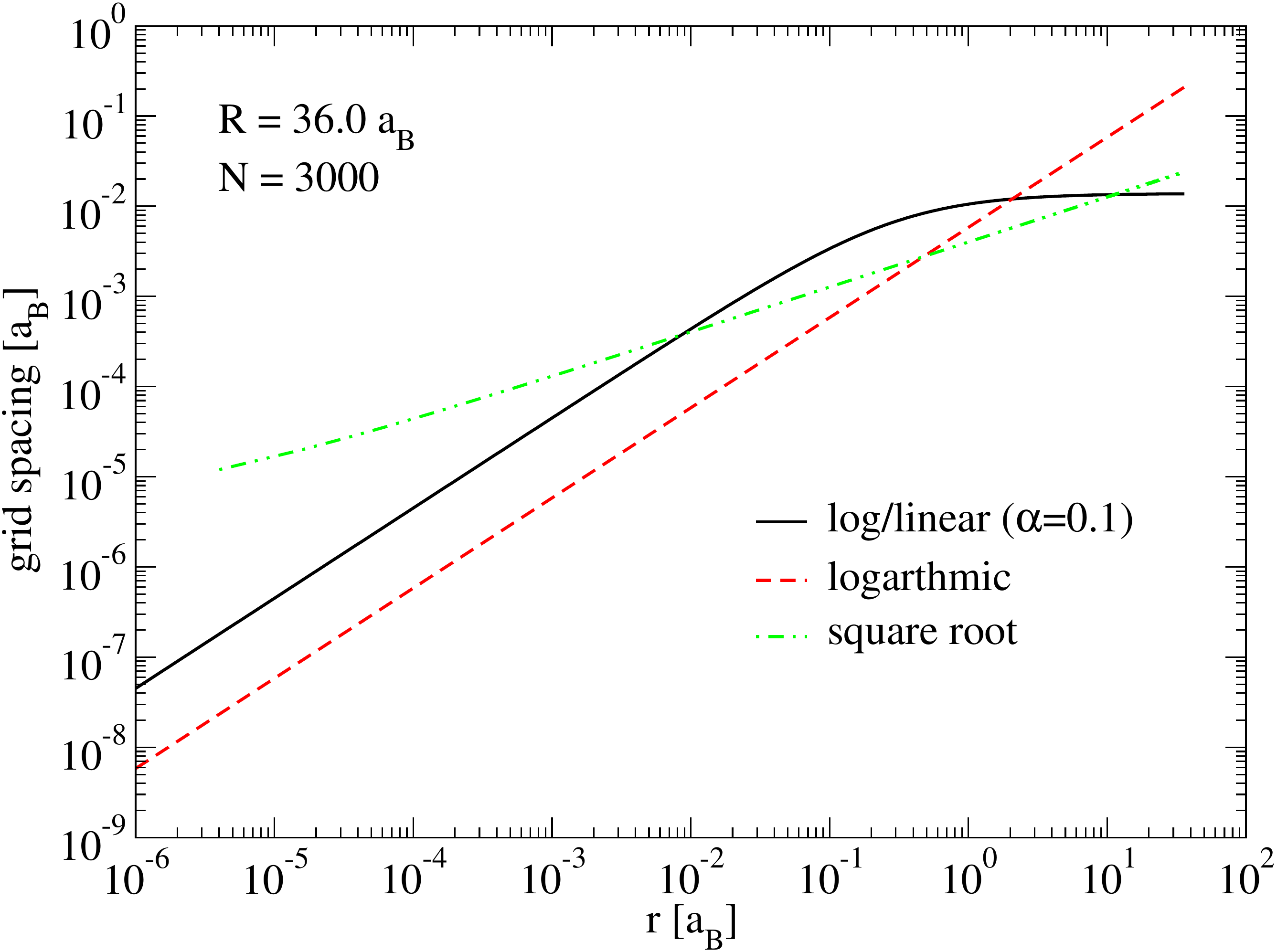}
\end{center}
\caption{(Color online) 
  Grid spacing $r_{i+1} - r_{i}$ for various grid generation methods described in the text.
}
\label{fig_grid}
\end{figure}
For the radial grid we have tried one based on $\sqrt{r}$.  A disadvantage is that this does not allow one to vary the total
number of grid points $N$ independently from the value of $r$ at the first grid point $r_1$.  Since the value of $r$ at
the end of the grid is fixed by the ion-sphere radius $R$, then $r_1 = R/N^2$.  This lack of flexibility is problematic.  We
have also tried a grid based on $\log r$ \cite{ondrej_bnd}, which allows such flexibility, but for low densities requires many grid points
to maintain resolution near the sphere boundary.  Finally, we settled on the log-linear grid presented in \cite{wilson11}.
This is logarithmic near the origin and so has enough points to resolve wavefunctions which can vary rapidly for small $r$,
and switches to linear spacing as $r$ increases.  We have found this grid to be generally accurate from low to high density, and from
low $Z$ to high $Z$.  We have found $r_1 = 1.0\times 10^{-6}\, a_B$ and $N=3000$ to be robust for the applications presented here.  The log-linear
grid also requires a parameter $\alpha$ to be chosen which determines how quickly it switches from logarithmic to linear.
We have found $\alpha = 0.1 $ to be generally reasonable.  Note that $\alpha = 0$ corresponds to a purely
logarithmic (exponential) grid.  

Examples of the grid spacing from these three grid choices are shown in figure \ref{fig_grid}.  For this case
(lutetium at 0.01 g/cm$^3$, grid independent of temperatures) we find that $r_1$ and the grid spacing for the $\sqrt{r}$ grid are too large for
accurate convergence.  We also find that the $\log\,r$ grid is too sparse for large $r$.  Only the log-linear
grid has the resolution everywhere that is needed.  Note that we have implemented the Adams method so that \texttt{Tartarus}
can use any grid provided that $r$ can be transformed onto a linearly spaced grid $x$ and $dr/dx$ is smooth and
can be calculated \cite{johnson_book,ondrej_bnd}.

\subsection{Contour Integrals for Green's Functions}
The main advantage of using the Green's function is that it is analytic in
the complex plane, allowing energy integrals along the real energy axis
to be deformed to complex energy $z$ using Cauchy's integral theorem.  The
electron density can be calculated thus
\begin{eqnarray}
\label{gf_dens_cx}
n_{e}(r) & = & \frac{1}{\pi}\Im \int_{C}dz f(z,\mu)TrG(r,z) \nonumber\\
            && \!\!\!\!\! + 2k_{B}T\Re \left\{
              \sum_{j=1}^{N_{mat}}TrG(r,z_{j})\right\} 
\end{eqnarray}
$C$ refers to a contour that closes when joined to the real axis \cite{starrett15}, and the
sum over $j$ is a sum over the $N_{mat}$ poles (known as Matsubara poles) of the Fermi-Dirac function enclosed by
this closed contour, at energies $z_j = \mu + \imath \pi (2j-1) k_B T$.  Similarly, for equation of state calculation
we can use
\begin{eqnarray}
\label{gf_ke_cx}
\!\!\!\!\! \int_{-\infty}^{\infty}d\epsilon f(\epsilon,\mu)\chi(\epsilon)\epsilon
&\!\!\!\!\! = \!\!\!\!\!& \!\!\! \frac{1}{\pi}\Im \int_{V^{ion}}\!\!\!\! d^3r\!\!
                          \int_{C}dz f(z,\mu)\, z\, TrG(r,z) \nonumber\\
            && \!\!\!\!\!\!\!\!\!\!\!\!\!\!\!\!\!\!\!\!\!\!\!\! 
               \!\!\!\!\!\!\!\!\!\!\!\!\!\!\!\!\!\!\!\!\!\!\!\! 
            + 2k_{B}T\Re \left\{\int_{V^{ion}}d^3r\,\sum_{j=1}^{N_{mat}}\, z_j\, TrG(r,z_{j})\right\} 
\end{eqnarray}
The advantage of carrying out the energy integrals in the complex plane is twofold: 1) sharp
features in the integrand that occur for real energies are broadened in the complex
plane.  Hence resonances in the positive energy states that need to be tracked and
resolved on the real energy axis are broad and smoothly varying in the complex plane.  This
point was explored in detail in \cite{starrett15} (also see figure \ref{dos_1}), and 2) negative energy (or bound) states
are treated in exactly the same way as positive energy states.  The search for bound
states is tricky to make generally robust, and states with very small energies 
(eg. $|\epsilon| < 1\times 10^{-4} E_h$) can be especially hard to accurately represent.
By designing the contour $C$ so that it returns to the real axis at a large negative
energy, the search for bound states could be avoided altogether.  However, since deeply
bound states are very sparse in energy space, it makes sense to treat these more deeply 
bound states with the usual orbital approach, and the more weakly bound states
with Green's functions (see section \ref{sec_ne}).

We have used a rectangular contour, as in reference \cite{starrett15}.  To find $\epsilon_{min}$,
the energy at which the contour rejoins the real energy axis, we first solve for 
the bounds states using standard search methods (eg. reference \cite{johnson_book}), and look for
the highest lying (least negative) energy gap $\ge$ 10 E$_h$ between two states in an energy ordered list.  
$\epsilon_{min}$ is then set to be 1 E$_h$ less than the eigenenergy of the state on the higher lying side of
that gap.  This is done at each iteration of the SCF procedure to avoid double counting bound
states.
It may seem that since we are already finding the bound eigenstates we should just set $\epsilon_{min} = 0$
E$_h$.  In many cases this would work, but as mentioned above inaccuracies would occur if
our search algorithm missed bound states, or if bound states had very small energies.  The
Green's function approach avoids both of these pitfalls and as a result is very stable.

$\epsilon_{max}$ is set by requiring  $f(\epsilon_{max},\mu) \approx 10^{-10}$.  We split
the integration into panels and use a 4 point Gauss-Legrende scheme in each.  Care is taken
to resolve the Green's function near the Fermi-edge, which is important for highly
degenerate cases, i.e. when $k_B\,T / E_F << 1$ ($E_F$ is the Fermi energy).  The total number of points used is
dependent on $\epsilon_{min}$, $\epsilon_{max}$ and $\mu$, but typical values are 1000 to 2000
energy points.

While any contour can be used to carry out the energy integrals above, calculation
of the entropy is special. Due to the many valued logarithm in (\ref{fs}) the contour
cannot pass the branch-cut parallel to the first Matsubara pole at $\Im z = \pi k_B T$.
For sufficiently high temperature $\pi k_B T $ is greater than the imaginary part of the energy anywhere on the
contour and so the SCF contour can be used for $S$.  Typically this is so for $k_B T \gtrsim$ 10 eV.  For temperatures less than this
we have decided to use a purely orbital based density of states calculation, only for the
entropy at the end of the SCF procedure.  Hence we use all bound orbitals and
a resonance tracker \cite{wilson06} for the positive energy states.  Fortunately,
at such relatively low temperatures resonance tracking is less challenging and
we have found this to be accurate enough for $S$.  Note that even for these low $T$ cases
Green's functions are still used in the SCF procedure where we find they offer enhanced stability.

\subsection{Density Construction\label{sec_ne}}
Electron density $n_e(r)$ is the key quantity in density functional theory and
must be constructed accurately.  While it is in principle possible to construct $n_e(r)$
directly from the Green's function or the orbitals, it is very difficult to do this
robustly over a wide range of temperatures or densities and materials.  Instead
we have used the following numerically advantageous hybrid decomposition 
\begin{equation}
\begin{split}
n_e(r) = & n_e^{\sss core}(r)
       + n_e^{\sss GF}(r, l_{max})
       + n_e^{\sss ctm}(r, l_{max})\\
       &- n_e^{\sss free}(r, l_{max})
       + n_e^0
\end{split}
\end{equation}
Here $n_e^{core}(r)$ is the density due to bound states with $\epsilon < \epsilon_{min}$
(from equation (\ref{ne_dir_ss}))
\begin{equation}
  n_e^{\sss core}(r)   = \sum\limits_{i\in B, \epsilon_i<\epsilon_{min}} f(\epsilon_i,\mu) \frac{ 2|\kappa_i| } {4\pi r^2} 
            [P^{2}_{\kappa_i}(r,\epsilon_i) + Q^{2}_{\kappa_i}(r,\epsilon_i)]
\end{equation}
$n_e^{\sss GF}(r,l_{max})$ is calculated using (from equation (\ref{ne_gf_aa}))
\begin{equation}
n_e(r) = -\frac{1}{\pi}\Im \int_{\epsilon_{min}}^{\epsilon_{max}}d\epsilon f(\epsilon,\mu)TrG(r,\epsilon, l_{max})
\end{equation}
with $TrG(r,\epsilon,l_{max})$ calculated using (from equation (\ref{gf_rel_ss}))
\begin{eqnarray}
TrG(r,\epsilon, l_{max}) & = & - \imath p(1+\frac{\epsilon}{2mc^{2}}) 2m \nonumber\\
              &&  \!\!\!\!\!\!\!\!\!\!\!\!\!\!\!\!\!\!\!\!\!\!\!\!\!\!\!\!\!\!\!\!\!\!\!\!\!\!\!\!
              \!\!\!\!\!\!\!\!\!\!\!\!\!\!\!\!\!\!\!
              \times
              \sum_{l=0}^{l_{max}} \sum_{\kappa}
              \frac{2|\kappa|}{4\pi r^2}
                 [P_{\kappa}^{R}(r,\epsilon)P_{\kappa}^{I}(r,\epsilon) + Q_{\kappa}^{R}(r,\epsilon)Q_{\kappa}^{I}(r,\epsilon)]
\end{eqnarray}
where the sum over $\kappa$ runs over the allowed values of $\kappa$ for a given $l$, i.e.
for $l>0$, $\kappa = \{l,-l-1\}$ and for $l=0$, $\kappa = -1$.

$n_e^{\sss ctm}(r,l_{max})$ is the density given by (from equation (\ref{ne_dir_ss}))
\begin{equation}
\begin{split}
  n_e^{\sss ctm}(r,l_{max})  = &
             \int_{0}^{\epsilon_{max}}d\epsilon f(\epsilon,\mu)
              \sum_{l=l_{max}+1}^{l_{con}} \sum_{\kappa} \\
&  \times    \frac{ 2|\kappa| } {4\pi r^2} 
            [P^{2}_{\kappa}(r,\epsilon) + Q^{2}_{\kappa}(r,\epsilon)]
\end{split}
\end{equation}
$n_e^{\sss free}(r)$ is given by
\begin{equation}
\begin{split}
  n_e^{\sss free}(r)  = &
             \int_{0}^{\epsilon_{max}}d\epsilon f(\epsilon,\mu)
              \sum_{l=0}^{l_{con}} \sum_{\kappa}\\
&  \times    \frac{ 2|\kappa| } {4\pi r^2} 
            [{P^{0}_{\kappa}}^2(r,\epsilon) + {Q^0_{\kappa}}^2(r,\epsilon)]
\end{split}
\end{equation}
where the superscript $0$ on the orbitals indicates solution to the
Dirac equation with $V^{eff}(r) = 0$, i.e. the ``$free$'' solution.
$l_{con}$ is determined \cite{blenski} by incrementing $l$ and 
evaluating
\begin{equation}
\begin{split}
   &         \int_0^R dr\,\int_{0}^{\epsilon_{max}}d\epsilon f(\epsilon,\mu)
             \sum_{\kappa}
             2|\kappa|  \\
   & \times \left\{
            [P^{2}_{\kappa}(r,\epsilon) + Q^{2}_{\kappa}(r,\epsilon)]
            -
            [{P^{0}_{\kappa}}^2(r,\epsilon) + {Q^0_{\kappa}}^2(r,\epsilon)]
            \right\} \\
   & < \mbox{\texttt{TOL}}
           \label{lcon}
\end{split}
\end{equation}
until for two consecutive $l$'s this condition is true.  We have found \texttt{TOL} $=10^{-4}$ to be
robust.  $n_e^0$ is the free electron gas density for temperature $T$ and chemical potential $\mu$.  It is
used to correct the electron density that has had $n_e^{free}(r)$ removed
and the $l$ sum truncated at $l_{con}$
\begin{equation}
\begin{split}
  &n_e^0  =  
             \int_{0}^{\infty}d\epsilon f(\epsilon,\mu)
              \sum_{l=0}^{\infty} \sum_{\kappa}
            \frac{ 2|\kappa| } {4\pi r^2} 
            [{P^{0}_{\kappa}}^2(r,\epsilon) + {Q^0_{\kappa}}^2(r,\epsilon)] \\
           & =c_{\sss TF} \left[ F_{\frac{1}{2}}(\mu/k_BT, k_B T/mc^2)
           + \frac{k_B T}{mc^2} F_{\frac{3}{2}}(\mu/k_BT, k_B T/mc^2)
           \right]
           \label{ne0}
\end{split}
\end{equation}
where $c_{\sss TF} \equiv  \sqrt{2}(k_B T)^{\frac{3}{2}} / \pi^2$ and
\begin{equation}
F_{n}(\eta,\beta)=\int_{0}^{\infty}dx\frac{x^{n}\sqrt{1+\hf\beta x}}{e^{x-\eta}+1}
\end{equation}
are the relativistic Fermi-Dirac integrals \cite{Gong}.
Hence electrons in states with $l>l_{con}$ are treated as free electrons.  The
convergence of equation (\ref{lcon}) ensures that this approximation is accurate.

$l_{max}$ controls which states are treated with Green's functions, and which are
treated with orbitals.  Typically we choose $l_{max} \approx 40$, which ensures
any resonances in these angular momentum channels are correctly integrated.  For
$n_e^{\sss ctm}(r)$ and $n_e^{\sss free}(r)$ we use a fixed energy grid, based on
a linearly spaced $\sqrt{\epsilon}$ grid and typically use 400 points.  Using orbitals on this fixed 
energy grid is very rapid, more so than the Green's function evaluation which uses
a denser energy grid.  Moreover the Green's function requires both the regular and 
irregular solutions, whereas the orbital only requires one solution of the Dirac equation.
The above decomposition is robust for the cases studied here.  The non-relativistic 
decomposition is identical and can be obtained from the above by setting $Q_\kappa =0$,
$k_B T / mc^2 = 0$ and $\epsilon/2mc^2 =0$.

This decomposition scheme is also use to evaluate $U^k$ and $\mathcal{T}$.  Note
that, analgous to $n_e^0$, the free electron gas kinetic energy density $k_e^0$ is
\begin{equation}
\begin{split}
  k_e^0  =&  \int_{0}^{\infty}d\epsilon f(\epsilon,\mu)\chi^0(\epsilon)\epsilon \\
          & = c_{\sss TF} k_B T \left[ F_{\frac{3}{2}}(\mu/k_BT, k_B T/mc^2)\right. \\
          & + \left. \frac{k_B T}{mc^2} F_{\frac{5}{2}}(\mu/k_BT, k_B T/mc^2)
           \right]
\end{split}
\end{equation}
where $\chi^0(\epsilon)$ is the free electron density of states.  For $\mathcal{T}$
we have
\begin{equation}
\begin{split}
  k_e^{0,A}  =&  \int_{0}^{\infty}d\epsilon f(\epsilon,\mu){\chi^A}^0(\epsilon)\epsilon \\
          & = c_{\sss TF} k_B T \left[ F_{\frac{3}{2}}(\mu/k_BT, k_B T/mc^2) \right]
\end{split}
\end{equation}

One problem in solving for the Green's function is that at high $l$ the solution
near the origin becomes inaccurate because it results from the multiplication of a
very small regular solution and a diverging irregular solution.  
We have found that this does not present a problem for solution of the 
SCF problem where small $r$ dependence is suppressed with an $r^2$ from the
Jabobian.  However, for evaluation of the equation of state, integrals like 
$\int\,d^3r\, n_e(r) / r$ are required (eg. equation (\ref{fel})).  Hence
the result is more sensitive to the small $r$ behavior of $n_e(r)$.  Thus for
equation of state only, we have found it useful to replace $n_e(r)$ for
$r<10^{-4}$ a$_B$ with an orbital only calculation of the density.  Fortunately
since we only need the small $r$ part of this density and it is not needed
in the SCF calculation, it does not need to be highly accurate.  Hence
we use a purely orbital based calculation for this part of the
density for such integrals only.  As for entropy at low temperature, we use all core orbitals
and a resonance tracker to replace the Green's function calculation.

\subsection{Self Consistent Field Acceleration\label{sec_scf}}
\begin{figure}
\begin{center}
\includegraphics[scale=0.33]{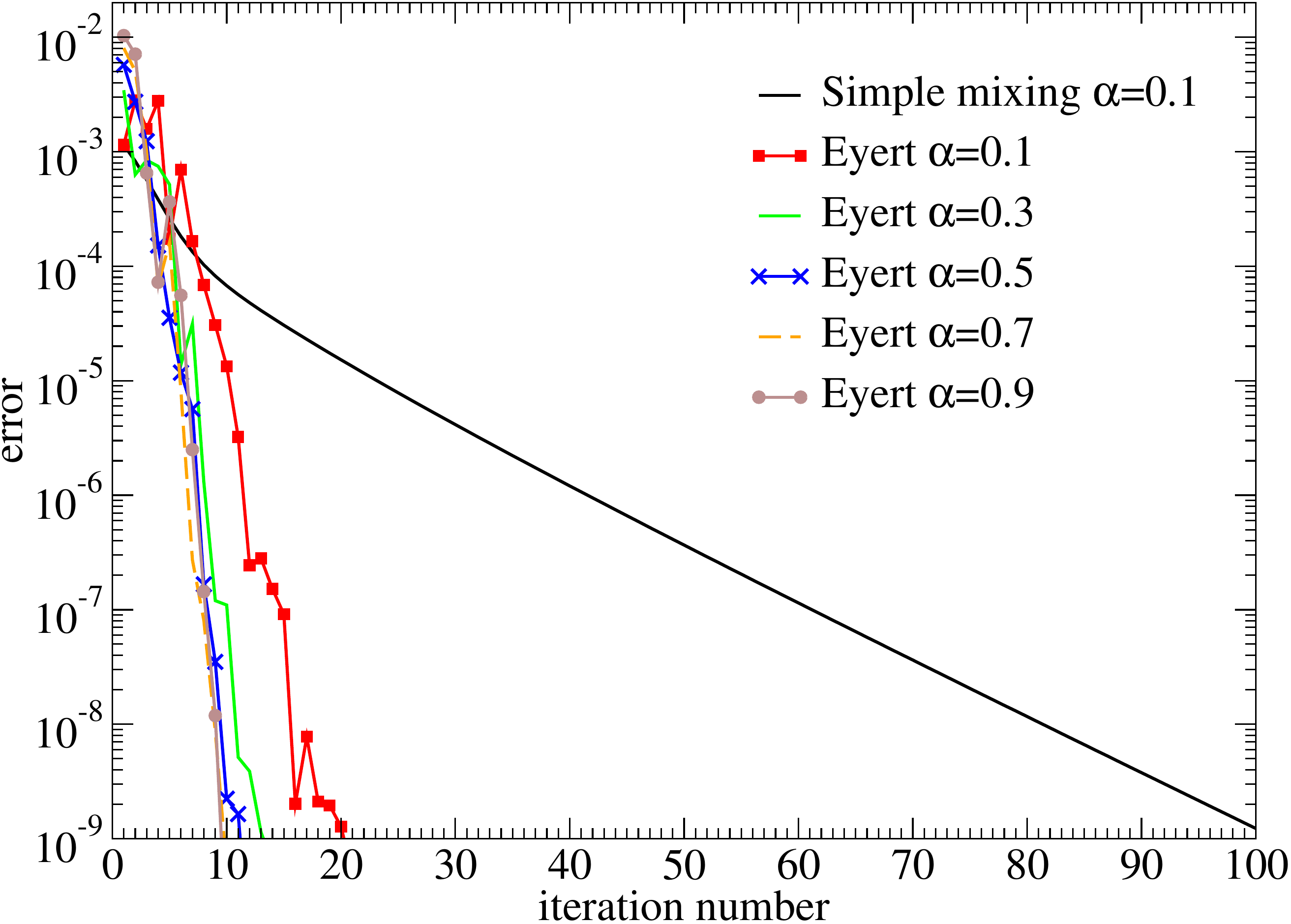}
\end{center}
\caption{(Color online) 
  Example of SCF acceleration for lutetium at 10 eV and 10 g/cm$^3$.  We compare simple mixing
  to Eyerts method with $M=5$, as a function of the mixing parameter $\alpha$.
}
\label{scf_1}
\end{figure}
Let us denote as $|x\rangle$ a vector generated from $V^{eff}(r)$ or equivalently $n_e(r)$, where
the components of the vector correspond to the grid points in no particular order.
The SCF procedure is 
\begin{enumerate}
\item
Begin with an initial guess of $|x\rangle$.
\item
Generate output vector $|x^{out}\rangle$.  For example, if $|x\rangle$ is $V^{eff}(r)$ we would solve the Dirac
equation, generate $n_e(r)$, and then calculate an output potential $V^{eff,out}(r) = |x^{out}\rangle$ by solving the Poisson 
equation and adding the exchange and correlation potential.
\item
Calculate $|F\rangle = |x^{out}\rangle - |x\rangle $.
\item
SCF convergence is achieved when $|F\rangle = |0\rangle$.  If not achieved, generate new $|x\rangle$ and return to step 2.
\end{enumerate}
\begin{figure}
\begin{center}
\includegraphics[scale=0.33]{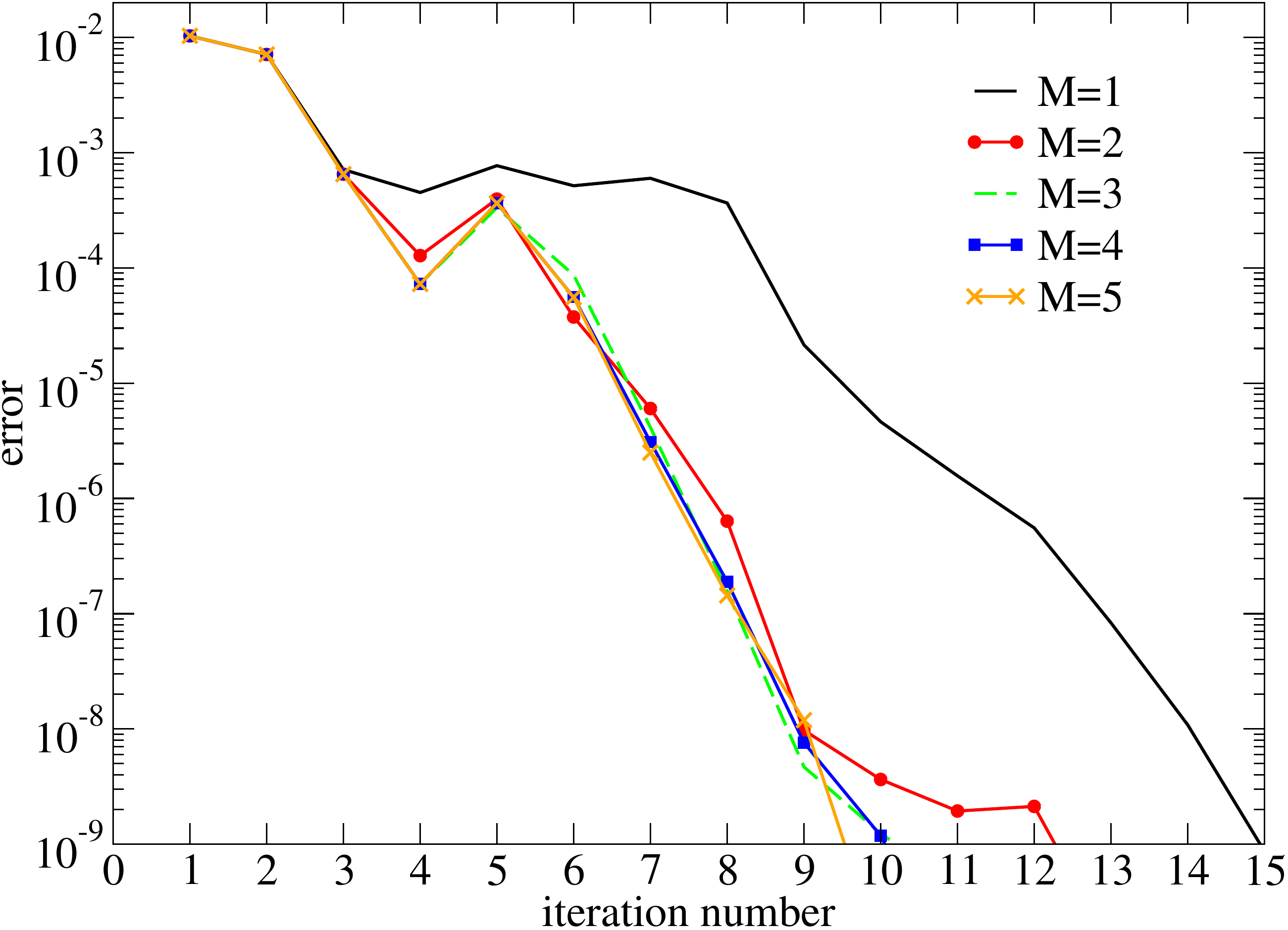}
\end{center}
\caption{(Color online) 
  Example of SCF acceleration for lutetium at 10 eV and 10 g/cm$^3$.  We compare 
  Eyerts method with $\alpha = 0.9$ as a function of the order $M$.
}
\label{scf_2}
\end{figure}

In the frequently used simple mixing method the new $|x\rangle$ in step 4 is generated with
\begin{equation}
|x^{(m+1)}\rangle = |x^{(m)}\rangle + \alpha |F^{(m)}\rangle
\end{equation}
where $m$ labels the SCF iteration number.  $\alpha$ is a mixing parameter that can be adaptive, or fixed.
Typically a small value $\alpha = 0.1$ is needed for robust convergence and perhaps 80-100 iterations is
necessary.
A much more robust scheme that greatly reduces the number of iterations required to reach convergence
has been given in the work of Eyert \cite{eyert96}.  To our knowledge this has not been explored for 
average atom models before.  Eyert's work is a correction and extension of the more famous Anderson mixing scheme \cite{anderson65}.
In this scheme $|x\rangle$ is generated with
\begin{equation}
|x^{(l+1)}\rangle = |x^{(l)}\rangle + \alpha |F^{(l)}\rangle - \sum_{m=l-M}^{l-1} \gamma_m^{(l)} 
\left[
|\Delta x^{(m)}\rangle +\alpha |\Delta F^{(m)}\rangle
\right]
\label{eand}
\end{equation}
where $M$ is the order of the mixing (an input choice), the $\gamma_m^{(l)}$ are coefficients to be determined, and  
\begin{eqnarray}
|\Delta x^{(m)}\rangle  & = & |x^{(m+1)}\rangle - |x^{(m)}\rangle \\
|\Delta F^{(m)}\rangle  & = & |F^{(m+1)}\rangle - |F^{(m)}\rangle
\end{eqnarray}
For $M=0$ equation (\ref{eand}) recovers the simple mixing formula above.  For $M \ge 1$ we take into
account the input and output vectors from the previous $M$ iterations.
To find the coefficients $\gamma_m^{(l)}$ we solve a matrix equation
\begin{equation}
\underline{\gamma} = \underline{B}^{-1} \underline{A}
\end{equation}
where $\underline{\gamma} = [\gamma_m^{(l)}]$ is an $M \times 1$ matrix with $m=l-M,\ldots,l-1$, $\underline{A}$ is an $M \times 1$
matrix with elements $\langle\Delta F^{(n)}|F^{(l)}\rangle$ ($n=l-M,\ldots,l-1$). $\underline{B}$ is an $M \times M$ matrix with elements 
\footnote{The notation $\langle \Delta F^{(n)}|F^{(l)}\rangle$ means the inner product of the vectors.}
\begin{equation}
B_{nm} = (1 + w_0^2 \delta_{nm})\langle\Delta F^{(n)}| \Delta F^{(m)}\rangle 
\end{equation}
Note that $\underline{B}$ is a symmetric matrix.
Due to saturation of improvements for higher orders, $M$ is taken to be 5 or 6 at maximum \cite{eyert96}.  Hence the inversion
of the matrix $\underline{B}$ is rapid. $w_0^2$ is a small parameter that breaks the symmetry (and thus removes linear dependences in
Anderson's original method), it is fixed at $10^{-4}$.

Eyerts method is a quasi-Newton method.  It is equivalent to Broyden's method \cite{broyden65, johnson88} provided
certain choices are made in that method \cite{eyert96}.
The mixing parameter $\alpha$ for Eyert's method can be larger than for simple mixing.
In practice we set $|x\rangle = V^{eff}(r) \times r / Z$, and calculate an error using
the maximum value of the absolute value of $|F\rangle$.  We require error $<10^{-9}$
for two consecutive iterations.
In figure \ref{scf_1} we show an example of this, comparing the simple mixing method
with the safe choice of $\alpha=0.1$ to Eyert's method for various $\alpha$.  The
reduction in number of iterations, even with the same $\alpha$ is remarkable, and results
in a corresponding reduction in computational time.  Larger values of $\alpha$ lead to
improved errors, though the effect saturates by $\alpha=0.9$.  It is important to note that
not only is Eyert's method faster but it is also more stable than simple mixing, which can fail
to converge in certain cases requiring manual reduction of $\alpha$.  Indeed setting $\alpha=0.9$
and running this case with simple mixing the SCF loop fails to converge.  In figure \ref{scf_2} we show
the effect of the order of Eyert's method on the error.  The advantages saturate by
$M=5$.  Our default choice in \texttt{Tartarus} is $M=5$, $\alpha=0.9$.  We have found
this to be very stable, requiring no adjustment for any of the results presented here.

\begin{figure}
\begin{center}
\includegraphics[scale=0.33]{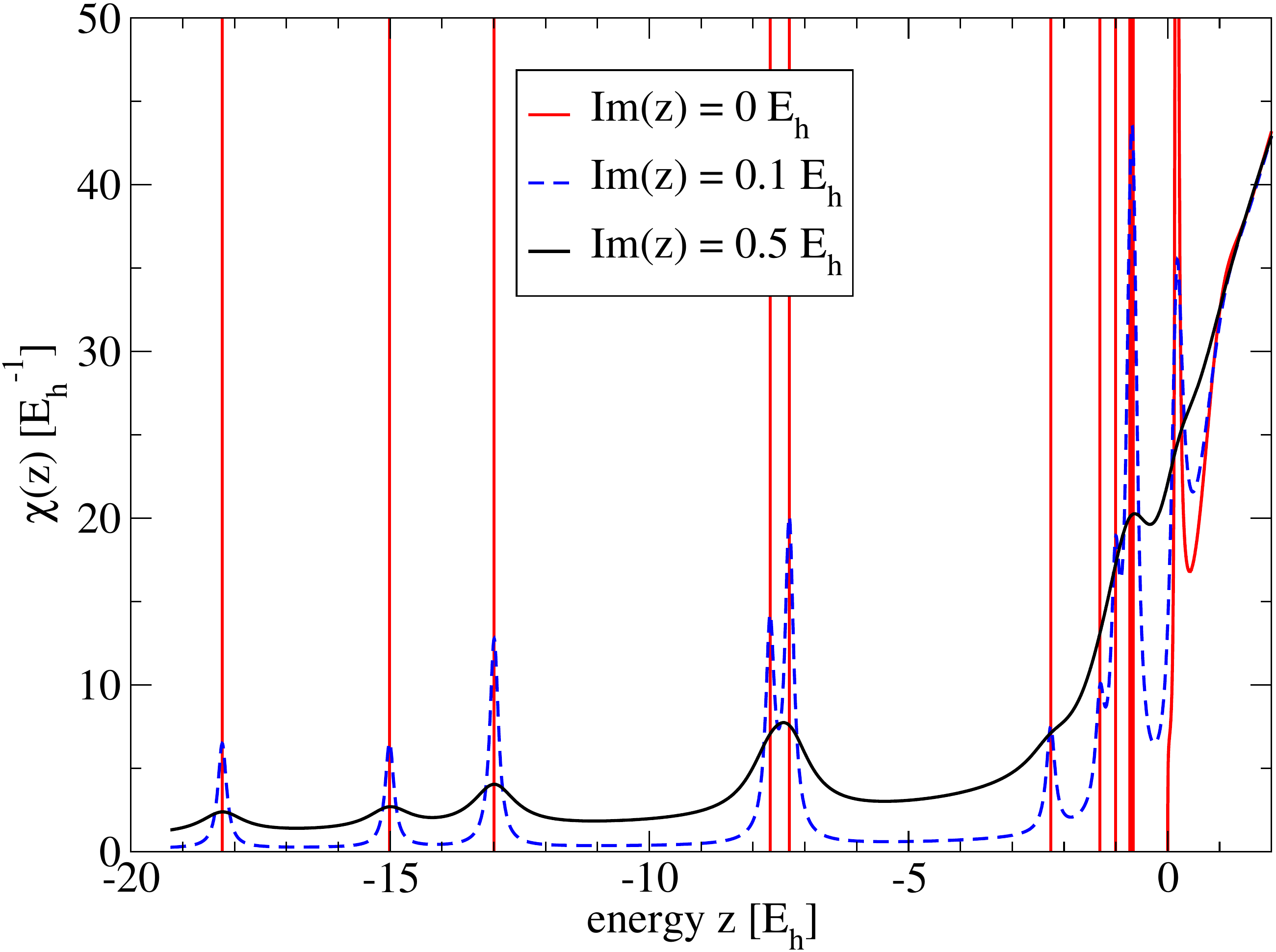}
\end{center}
\caption{(Color online) 
  Density of states $\chi(z)$ for lutetium at 10 g/cm$^3$ and 10 eV.  The solid
  red line is calculated using orbitals along the real energy axis.  Using Green's 
  functions we can evaluate $\chi(z)$ for complex energy $z$.  Increasing the imaginary
  part of $z$ features, including the discrete bound states and a continuum resonance, are
  broadened, making them easy to integrate over.  Note $\Im{z} = 0.5 $ E$_h$ is typical 
  for the horizontal part of our integration contour.
}
\label{dos_1}
\end{figure}
\begin{figure}
\begin{center}
\includegraphics[scale=0.40]{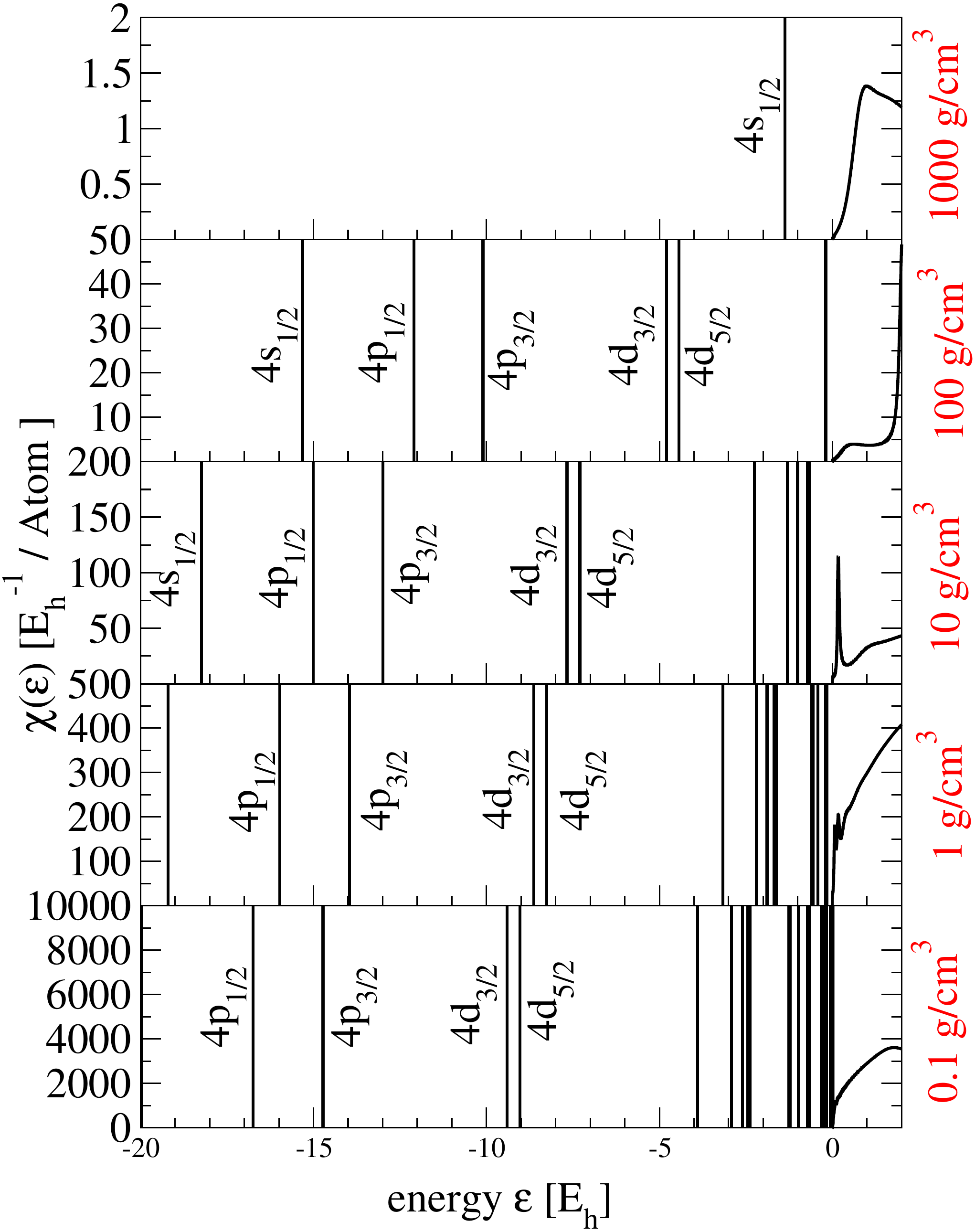}
\end{center}
\caption{(Color online) 
  Density of states $\chi(\epsilon)$ for lutetium at 10 eV for real energy $\epsilon$.  
}
\label{dos_2}
\end{figure}

\section{Example}
\subsection{Density of States}
In figure \ref{dos_1} the density of states $\chi(z)$ as a function
of complex energy $z$ is shown for lutetium at 10 eV and 10 g/cm$^3$.
For $\Im(z) = 0$ the calculation is purely in terms of orbitals.  We
used a bound state search algorithm, and the bound states appear in the
figure as vertical lines at negative energies, representing the
$\delta(\epsilon_i-\epsilon)$.  For positive energy states we used a resonance tracker, and
a resonance appears at $\sim 0.2$ E$_h$.  For $\Im(z) > 0$ the calculation
is purely in terms of Green's functions.  We see Lorentzian like line
shapes around each bound state energy and around the resonance.  For
$\Im(z) = 0.5$ E$_h$ the features are well smoothed out and integrating
over them is accurate and does not need adaptive mesh refinement, as a resonance
tracker does.  This is the principal advantage of using Green's functions.

In figure \ref{dos_2} the density of states $\chi(z)$ along
the 10 eV isotherm, from $\sim 1/100^{th}$ to 100 times
solid density is shown.  At the lowest density the most bound states exist (more
appear at more negative energies).  A few have been labeled in the figure to show that as
density increases the bound states move toward the continuum (positive energy) and
eventually disappear (pressure ionize).  A resonance appears if a state with $l>0$ is nearly bound.  On reducing
the density this resonance will transition to being a bound state with negative energy.
The resonance is a result of the centrifugal barrier term, $-l(l+1)/r^2$ in the
Schr\"odinger equation.  Hence there are no resonances associated with $l=0$
states.

\begin{figure}
\begin{center}
\includegraphics[scale=0.33]{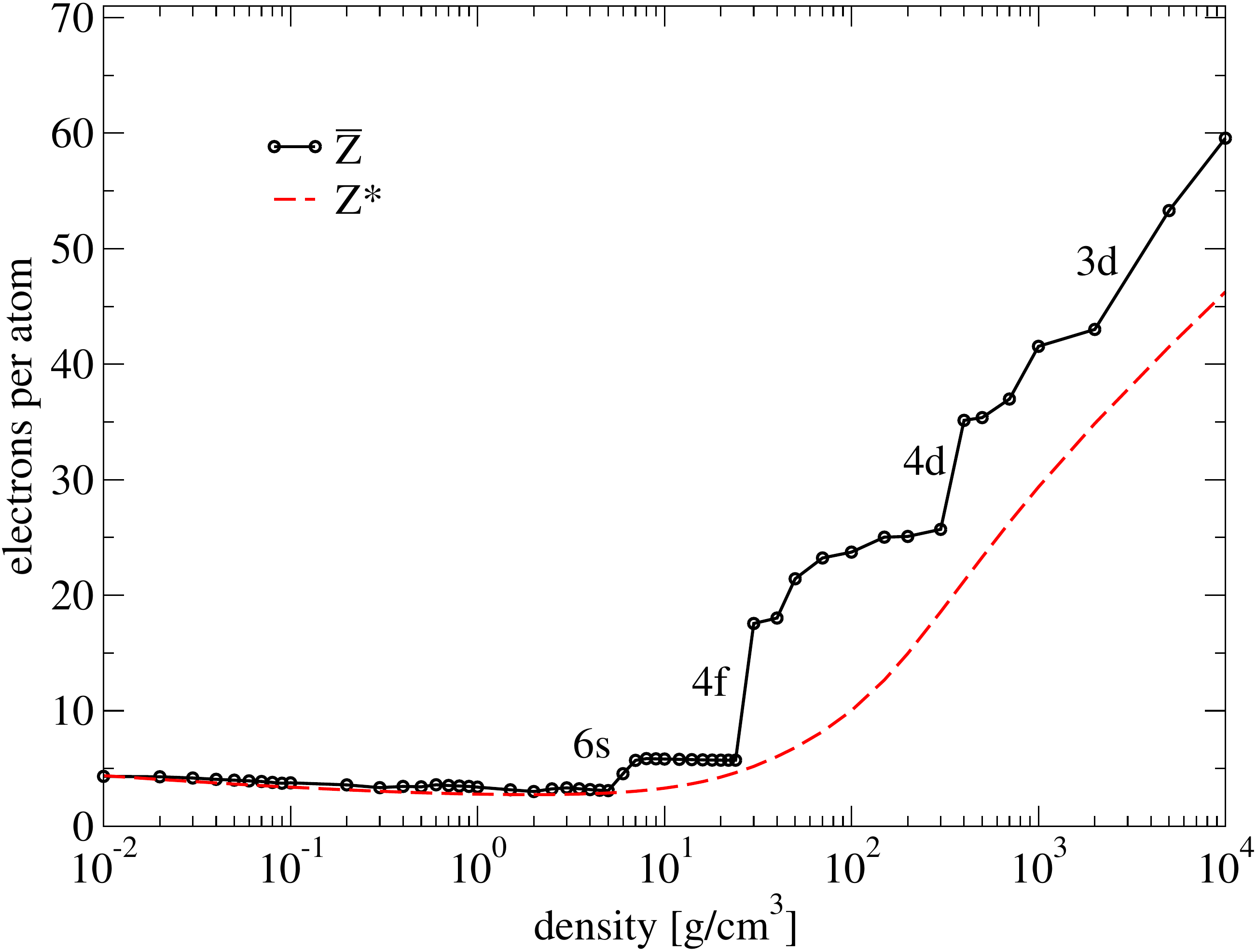}
\end{center}
\caption{(Color online) 
 Average ionization of lutetium at 10 eV.   Two definitions are explored.  Note that
 the definition choice does not affect in any way the properties of the average atom,
 for example, the equation of state does not depend on the definition.
 For $\bar{Z}$ the more prominent ionization features are labelled.
}
\label{ionization}
\end{figure}

\subsection{Extraction of Ionization}
A quantity of interest is the average ionization in the plasma.  This is
quantity is not uniquely definable, but given a definition it can be calculated from
\texttt{Tartatus}.  We stress that the ionization definition has no bearing on the model,
it does not influence in any way the results for the self-consistent solution or
the equation of state.  Here we explore two definitions.  The first is the number
of positive energy electrons $\bar{Z}$, defined as
\begin{eqnarray}
 \!\!\!\!\!\!\!\!\!\!\!\!\!  \!\!\!\!\!\!\!\!\!\!\!\!\!  \bar{Z}  & \!\!\!\!\!= \!\!\!\!\!& Z - \int_0^R dr\,\sum\limits_{i\in B} f(\epsilon_i,\mu)  2|\kappa_i|
            [P^{2}_{\kappa_i}(r,\epsilon_i) + Q^{2}_{\kappa_i}(r,\epsilon_i)] \nonumber\\
            && \!\!\!\!\!\!\!\!\!\!\!\!\!\!\!\!\!\!\!\!
            = \int_0^R dr\,\int_{0}^{\infty}d\epsilon f(\epsilon,\mu)
            \sum_{\stackrel{\kappa = -\infty}{\kappa \ne 0}}^{+\infty}
            2|\kappa|
            [P^{2}_{\kappa}(r,\epsilon) + Q^{2}_{\kappa}(r,\epsilon)]
\label{zbar}
\end{eqnarray}
The second definition is the number of free electrons per atom $Z^*$, i.e.
given $\mu$, $T$ and $V^{ion}$, the number of electrons per atom in a
free electron gas.  This is given by $Z^*=n_e^0 V^{ion}$, where $n_e^0$
is given by equation (\ref{ne0}).
The first definition $\bar{Z}$ has the benefit that it gives the expected ionization
in seemingly clear cut cases: for example $\bar{Z}=3$ for aluminum under normal conditions.
However it has the major disadvantage that it is generally discontinuous across
a pressure ionization.  When a state is ionized it ceases to be included in the bound state
sum, and instantly is counted in $\bar{Z}$.  In reality the ionized state retains
some of its bound like character if it appears as a resonance.  These meta-stable
resonance states are treated as fully ionized in the $\bar{Z}$ definition.  In figure
\ref{ionization} such discontinuities are observed for a lutetium 10 eV isotherm.
\begin{figure}
\begin{center}
\includegraphics[scale=0.33]{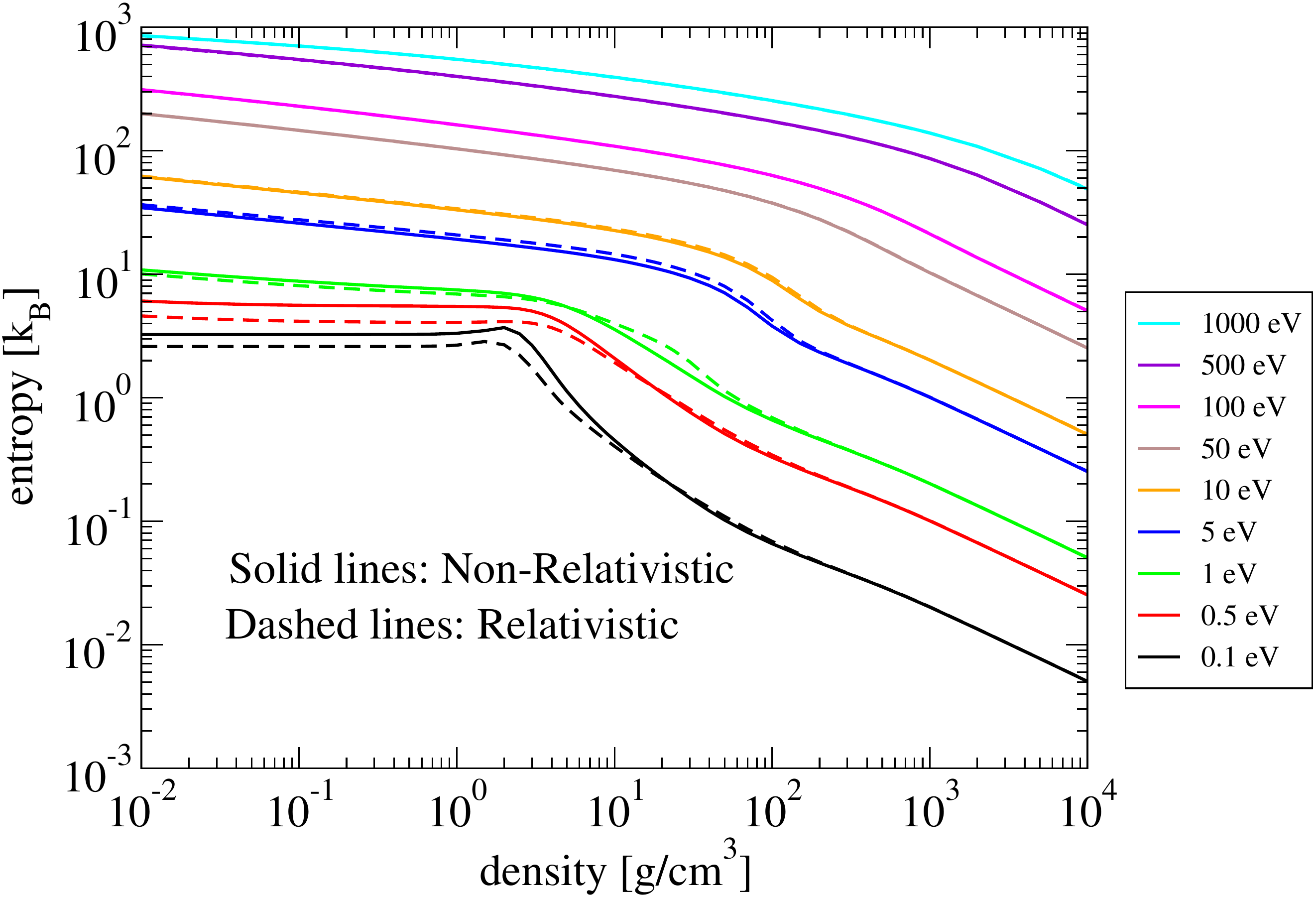}
\end{center}
\caption{(Color online) 
  Isotherms of entropy of lutetium plasma from \texttt{Tartarus}.  Both non-relativistic (solid lines)
  and relativistic (dashed lines) are shown for temperatures from 0.1 eV to 1 keV.  For any given
  density the entropy increases with temperature, as expected.
}
\label{lu_entropy}
\end{figure}

\begin{figure}[t]
\begin{center}
\includegraphics[scale=0.4]{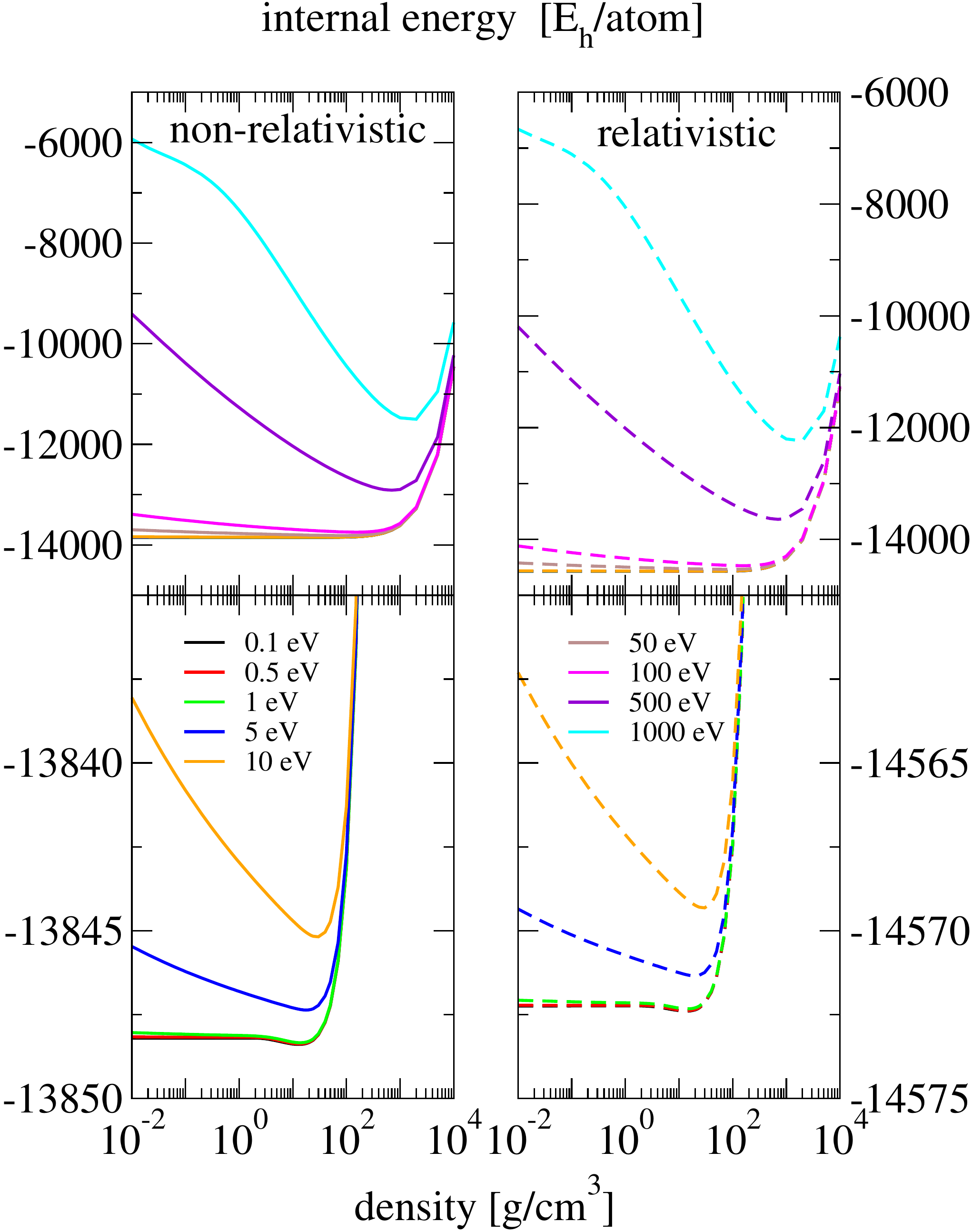}
\end{center}
\caption{(Color online) 
  Isotherms of internal energy of lutetium plasmas from \texttt{Tartarus}.  Both non-relativistic (solid lines)
  and relativistic (dashed lines) are shown for temperatures from 0.1 eV to 1 keV.  
}
\label{lu_energy}
\end{figure}

The second definition does not recover the expected ionization in cases like normal
density aluminum, where $Z^*\approx 2$.  However it is smooth across a pressure ionization
because the chemical potential $\mu$ is smooth, as it must be (figure \ref{ionization}).  Depending on the
application one can choose the definition that best suits.  But it is important to keep
in mind that the ionization depends on the definition.

\section{Case Study: Equation of State of Lutetium}
We now focus of an application of \texttt{Tartarus} to the equation of state of 
a high $Z$ material (lutetium, $Z$ = 71), from 0.1 to 1000 eV and $1/1000^{th}$ to
1000 times solid density ($\approx 10$ g/cm$^3$).
\begin{figure}[t]
\begin{center}
\includegraphics[scale=0.40]{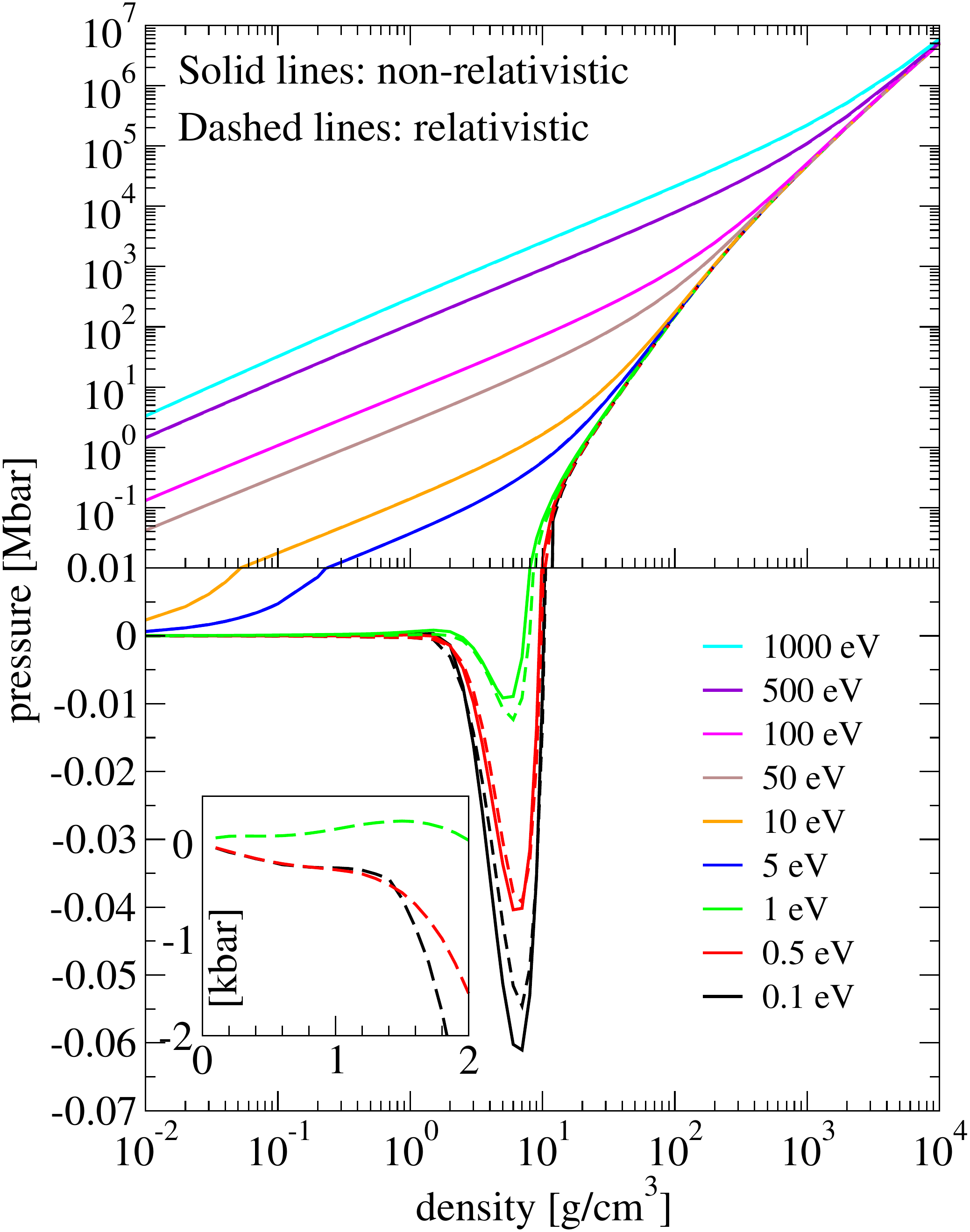}
\end{center}
\caption{(Color online) 
  Electron pressure of lutetium plasmas along isotherms.  Both relativistic and non-relativistic results
  are shown.
  The inset also shows electron pressure but focused on the low temperature region where entropy increases
  with density for fixed temperature.  The result is a region where pressure increases as temperature
  is lowered, for a fixed density.
}
\label{lu_pressure}
\end{figure}

In figure \ref{lu_entropy} entropy ($S$) isotherms are shown for both relativistic and non-relativistic
calculations.  For a given density $S$ increases with temperature, as expected.  For $k_B T \ge 1$ eV
$S$ always decreases as density is increased, again as expected.  However for $k_B T < 1$ eV there
is a region near normal density where the model predicts that $S$ increases with density.  This physically
unexpected behavior is not numerical inaccuracy but a consequence of the physical assumptions of the model \cite{sterne_talk}.
This behaviour is caused by the inconsistency between the normalization integral (\ref{norm}), which is over all
space, and the cell neutrality condition (\ref{neut}).  When a bound state has significant probability outside the
ion sphere, the number of electrons that bound state can contain becomes less that $2(2l+1)$  (non-relativistically).
The left-over electrons are forced into the positive energy states leading to an increase in $S$.  When
the temperature is high enough this effect still occurs but is overwhelmed by the entropy of the other 
ionized electrons.

The effect of relativity is generally modest, but it does make a significant difference at low temperatures
and densities.  This is because $S$ is dominated by the density of states near $\epsilon=\mu$ at low temperature.
For low densities the splitting of spin degeneracy in the Dirac equation results in the non-relativistic $5p$ state becoming
a $5p_{\frac{1}{2}} $ and $5p_{\frac{3}{2}}$, resulting in a change in the eigenvalue and therefore $\mu$.  For higher
densities, but still at low temperatures, $\mu>0$ and the splitting has a smaller effect since the eigenvalues are continuous.

For internal energy the results are shown in figure \ref{lu_energy}.  There is a significant
change in going from non-relativistic to relativistic due to significant relativistic
effects on the most tightly bound states.  At 0.1 eV, 10 g/cm$^3$ the eigenenergy of the $1s$ state changes from
$-2146.4$ E$_h$ to $-2318.8$ E$_h$.  Note that $|\epsilon_{1s}| / mc^2 \approx 0.12 $, so a significant
relativistic effect is expected.

In figure \ref{lu_pressure} the pressure due to electrons (i.e. no ideal ion contribution is added) is
shown.  In the top panel the relativistic and non-relativistic results are barely distinguishable on the
log-log scale.  The bottom panel shows the same data but on a linear pressure scale and focused on the
low pressure region.  
Approaching zero temperature at T=0.1 eV, Tartarus gives equilibrium volumes 10.1 g/cm$^3$ in the non-relativistic case 
and 10.3 g/cm$^3$ in the relativistic case, indicated by zero pressure. This is quite close to the room temperature crystal 
density of 9.84 g/cc. The negative pressure region is due to treating the material as a continuum, instead of as a mixed phase, 
such as a liquid-gas coexistence, which the model does not support.

\begin{figure}[t]
\begin{center}
\includegraphics[scale=0.33]{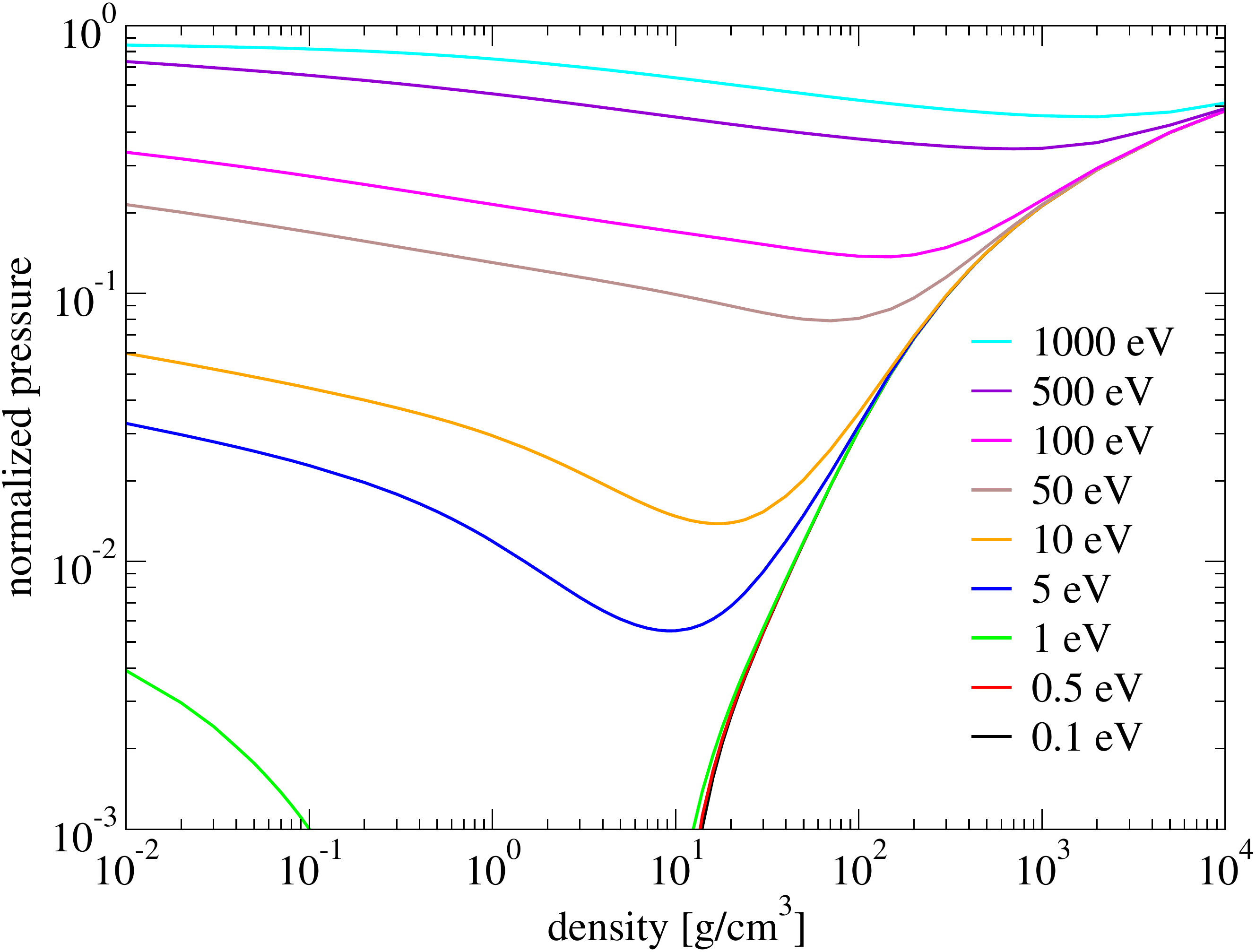}
\end{center}
\caption{(Color online) 
  Electron pressure for lutetium divided by the pressure of a non-interacting, relativistic, quantum electron gas
  with electron density $Z/V^{ion}$.
}
\label{lu_prel}
\end{figure}

In figure \ref{lu_prel} the electron pressure divided by that of a fully ionized ideal
electron system is shown.  The maximum value that this quantity can take is 1.  Even
at 10$^{-2}$ g/cm$^3$ and 1000 eV the normalized pressure is only $\approx 0.85$.  Under these
conditions the $1s_{1/2}$ and $2s_{1/2}$ states have eigenvalues of -2514.132 E$_h$ and -575.749 E$_h$
respectively and $\mu = $ -420.128 E$_h$, so that the Fermi-Dirac occupation factors are
1.000 and 0.986, i.e. nearly completely full.  Hence the reduction in pressure from the fully
ionized gas.

The Maxwell relation 
\begin{equation}
\left.\frac{\partial S}{\partial V}\right|_T
=
\left.\frac{\partial P}{\partial T}\right|_V
\end{equation}
\begin{figure}[t]
\begin{center}
\includegraphics[scale=0.33]{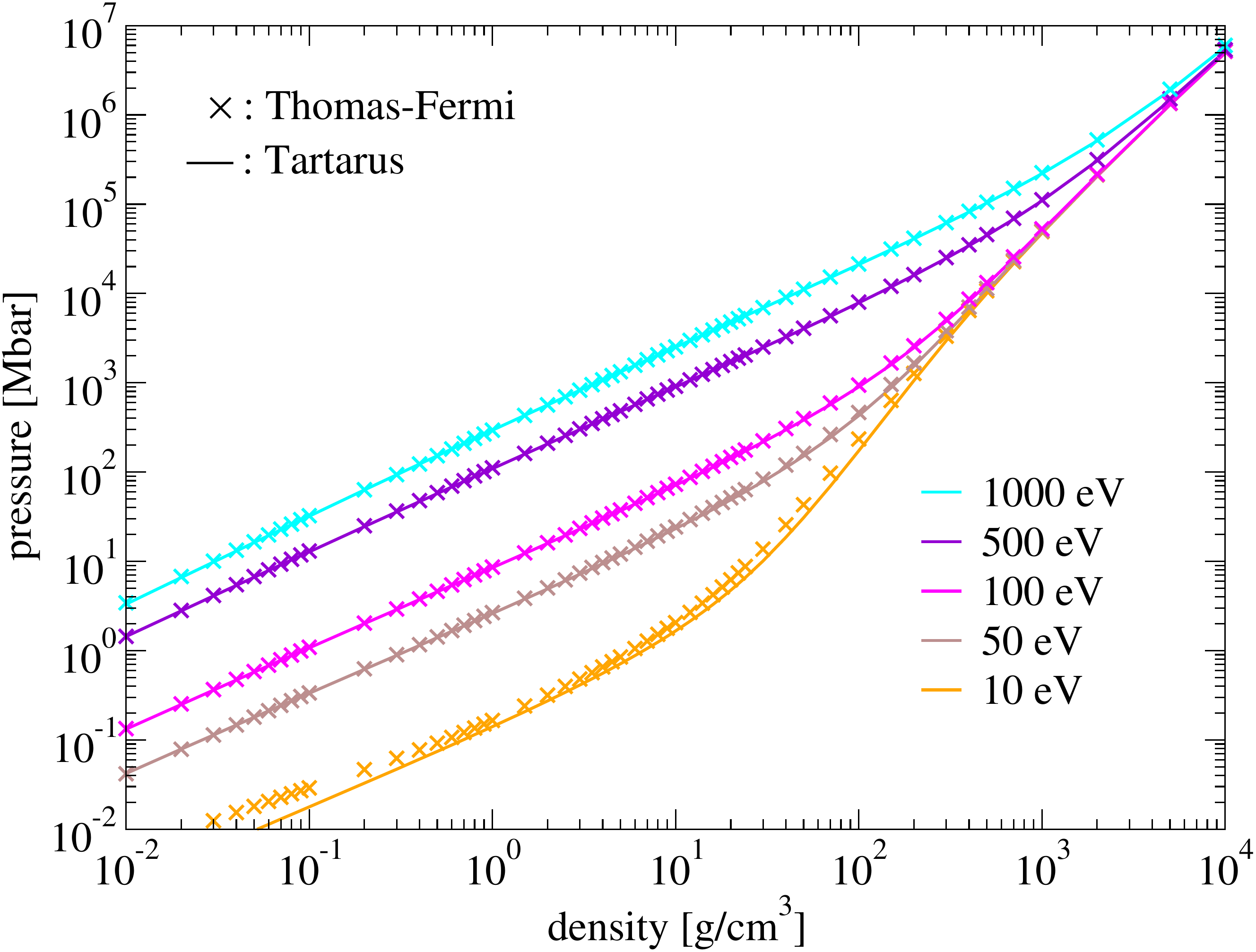}
\end{center}
\caption{(Color online) 
  Electron pressure for lutetium from \texttt{Tartarus} (relativistic) compared the Thomas-Fermi model's
  prediction \cite{feynman49}.  Note we have used the same exchange and correlation potential for
  both \cite{perdew81}.
}
\label{lu_tf}
\end{figure}
implies that the increase in $S$ with density for low temperatures observed in figure \ref{lu_entropy}
should correspond to a region where the pressure $P$ decreases as temperature increases, at constant
density.  In the inset in figure \ref{lu_pressure} such an effect is observed.  It is only seen for low temperatures
and only over the limited region in which $ \left.\frac{\partial S}{\partial V}\right|_T $ is negative.
Such a behavior is likely to be an artifact of the model.  This low temperature metal-to-nonmetal
transition region is difficult to model accurately and the present one-atom, spherically symmetric
model cannot be expected to fully capture this physics, though it clearly captures the gross effect.
\begin{figure}
\begin{center}
\includegraphics[scale=0.40]{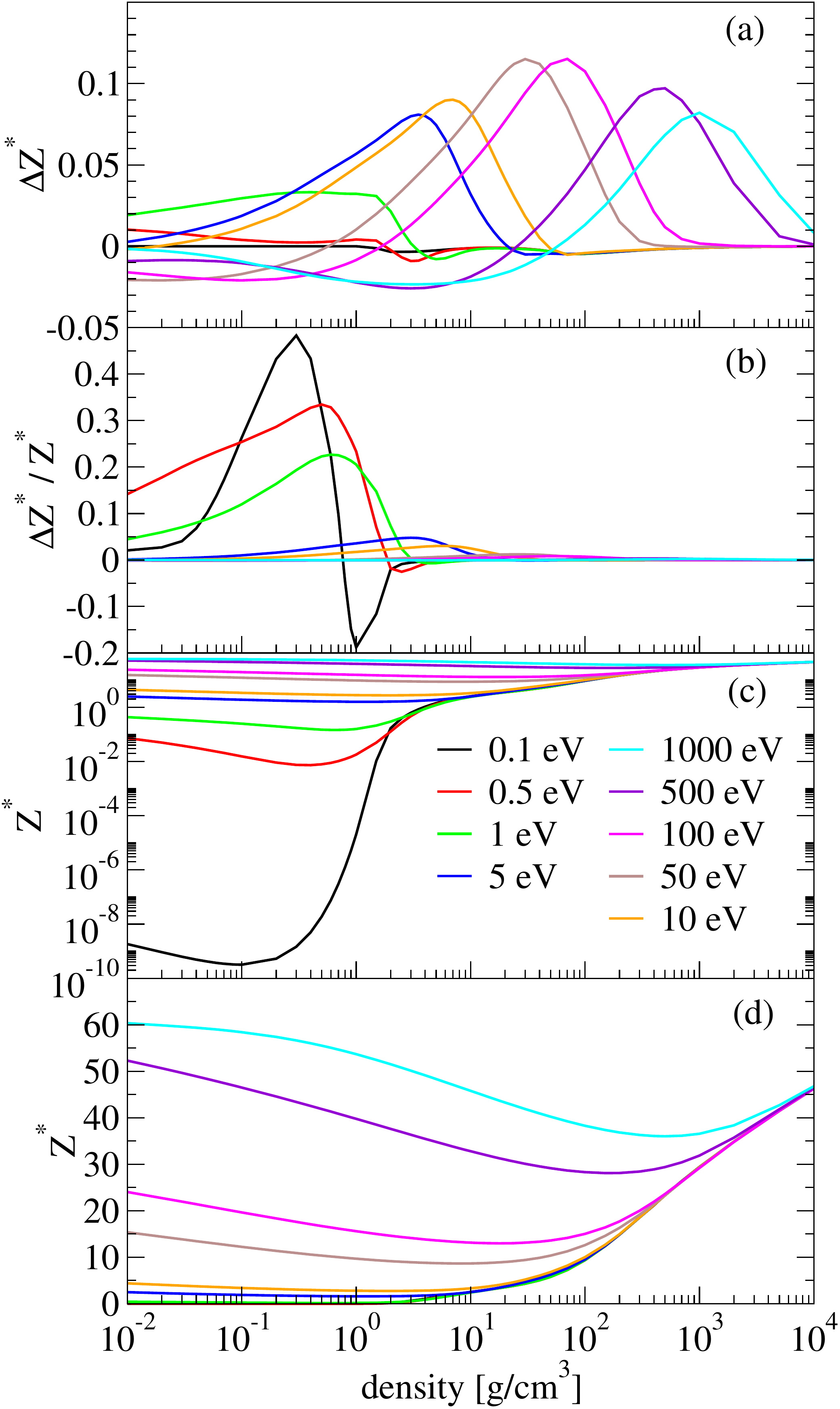}
\end{center}
\caption{(Color online) 
  The effect of finite temperature exchange and correlation on the average ionization per atom $Z^*$.
  We compare the zero temperature Perdew-Zunger (PZ) functional \cite{perdew81} to the recent finite temperature 
  functional \cite{groth17} (here labeled QMC17).
  In the panel (a) we show $\Delta Z^* = Z^*$(PZ) - $Z^*$(QMC17), in panel (b) we show $\Delta Z^* / Z^*$,
  where $Z^*$ is calculated with PZ.  The bottom two panels (c) and (d) show the same $Z^*$(PZ) but on different
  scales.
}
\label{lu_xc}
\end{figure}
\begin{figure}
\begin{center}
\includegraphics[scale=0.60]{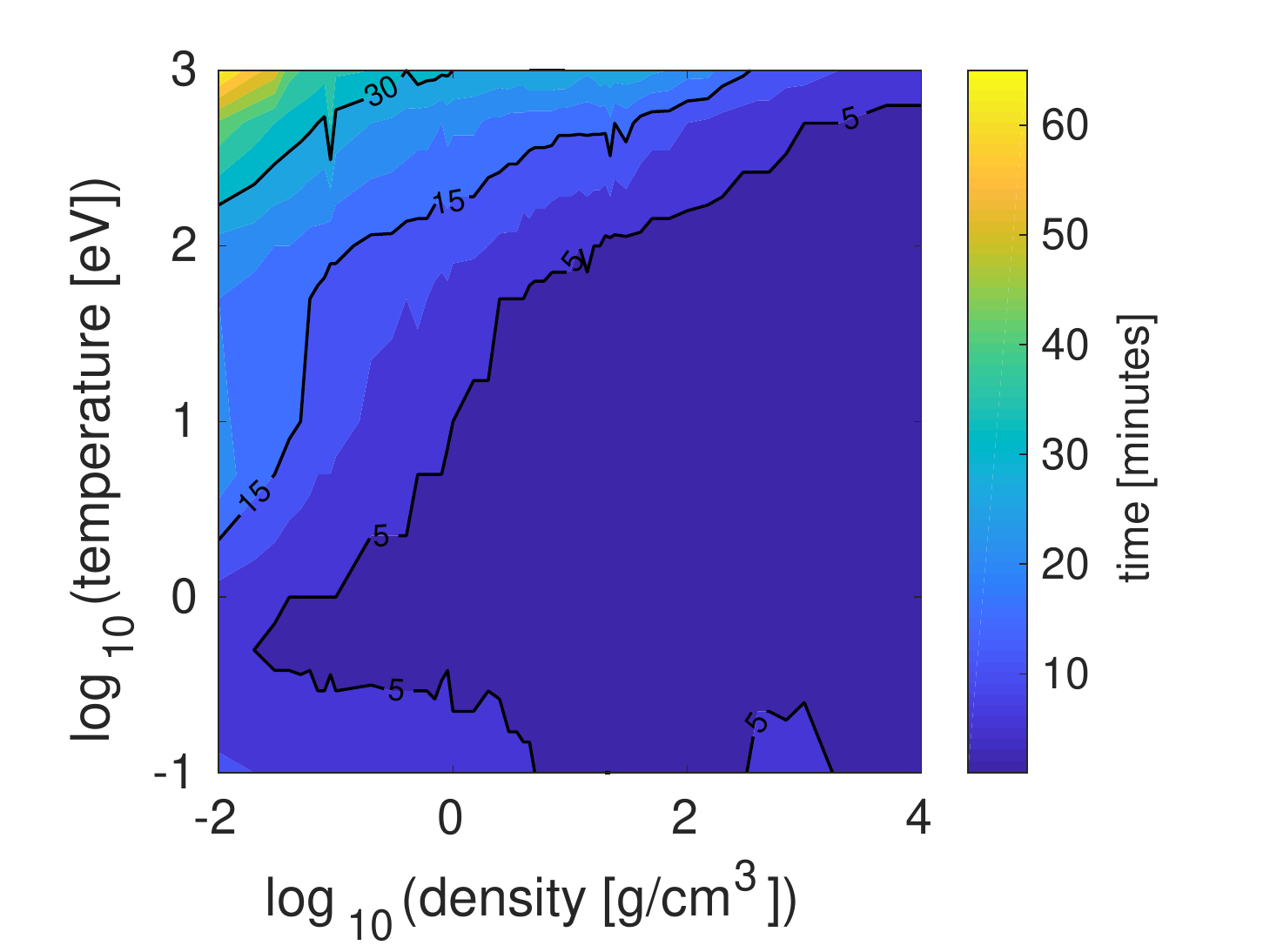}
\end{center}
\caption{(Color online) 
  Wall-time of {\texttt Tartarus} for the relativistic lutetium cases presented in figures \ref{lu_entropy} to
  \ref{lu_xc}.  Contour steps are at 5 minute intervals, and a select few have been labeled explictly. 
}
\label{lu_time}
\end{figure}

In figure \ref{lu_prel} we also observe a minimum in the normalized pressure.  This corresponds
to a minimum in ionization $Z^*$ (see figure \ref{lu_xc}).  Ionized electrons are the main
cause of electronic pressure \cite{blenski}.    The ionization increases with density for high densities
due to pressure ionization, a process analogous to the raising of energy levels in a square well potential
as the length of the square well is decreased.  Bound states disappear with increasing density and there are
insufficient bound states to hold all the electrons, so they are forced into positive energy states, i.e.
ionized.  At low densities, average ionization increases as density is lowered.  In this case
there are enough electron states to hold all the electrons but their Fermi-Dirac occupation factors
become $<$ 1.  This arises from the fact that the bound states approach their isolated atom limit, and
hence become insensitive to changes in density, however, the chemical potential continues to decrease,
leading to smaller Fermi-Dirac occupations factors for the same state.  The physical process
underlying this is photo ionization.  Though there are no radiation fields explicitly included
in the model's Hamiltonian, the assumption of local thermodynamical equilibrium implies that
the radiation temperature is equal to the electron temperature.   This is embedded in the 
Fermi-Dirac occupation factor, which does appear explicitly in the model.

\begin{figure}[t]
\begin{center}
\includegraphics[scale=0.40]{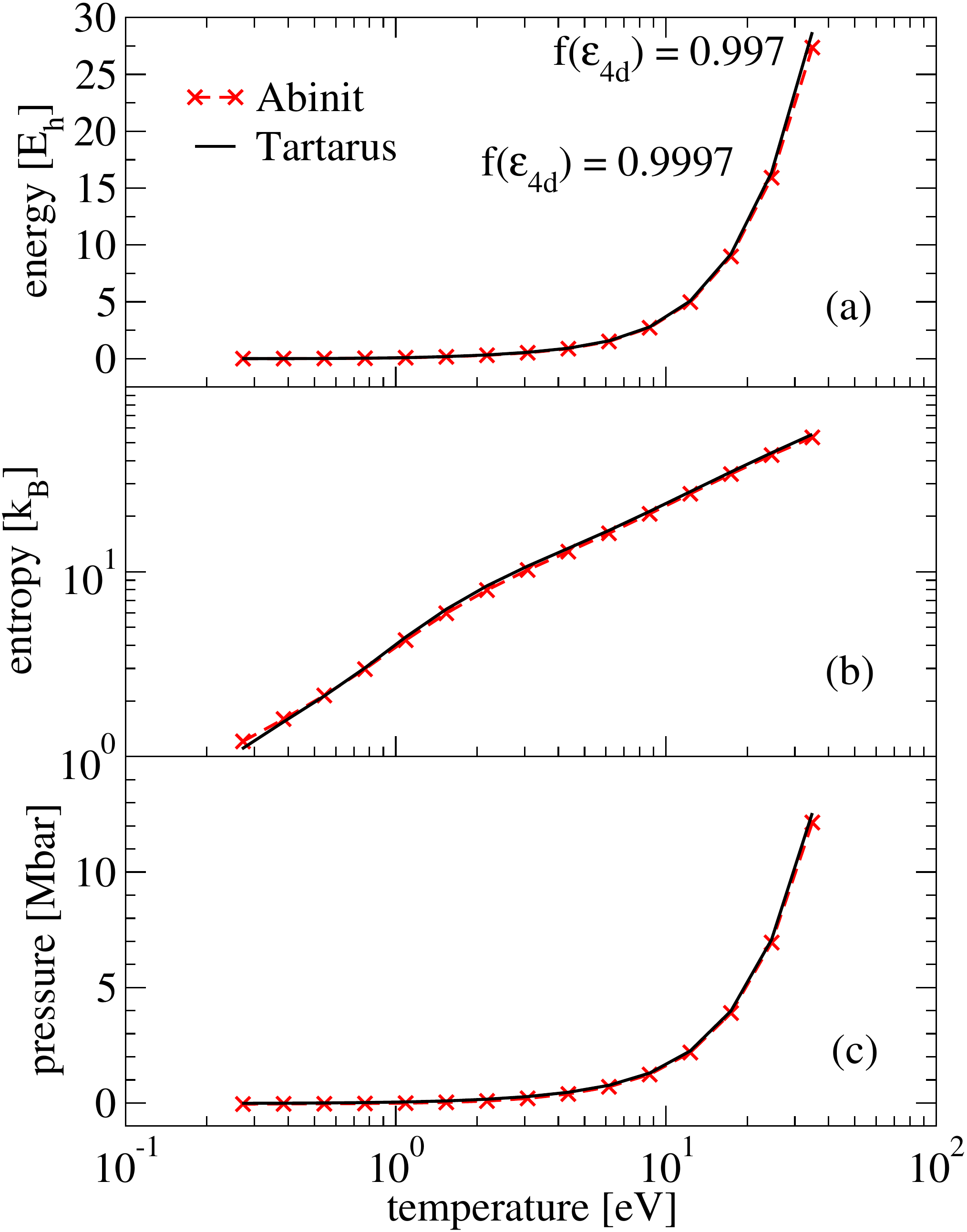}
\end{center}
\caption{(Color online) 
  Isocore (9.773 g/cm$^3$) of lutetium comparing results from the {\texttt Abinit} plane wave code using a
  HCP crystal structure to {\texttt Tartarus} results.  
}
\label{lu_isocore}
\end{figure}

In figure \ref{lu_tf} we compare the electron pressure from \texttt{Tartarus} to the generalized Thomas-Fermi (TF)
model \cite{feynman49}, using the same exchange and correlation potential \cite{perdew81}.  The TF model
is commonly used to construct equation of state tables \cite{sesame92,more88}, however it has a number of
well known drawbacks.  For example, it does not have shell structure and as a consequence its internal energy
is quite inaccurate.  However it is expected to give the correct pressure at high temperatures and densities.
In the figure we observe good agreement of the \texttt{Tartarus} electron pressure with the TF model for
high temperatures and densities, in line with this expectation.  Note that for truly free electrons the
two models become identical.  For the lowest temperature in the figure, 10 eV, significant deviations
between the models is seen due to the neglect of shell structure in the TF model.  The agreement 
between \texttt{Tartarus} and the TF model is a validation of 
our implementation in those limits.

All of the results so far presented have used a zero temperature local density approximation (LDA) 
exchange and correlation functional $F^{xc}$ \cite{perdew81}.  Recently, new temperature dependent LDA functionals
have become available \cite{karasiev14, groth17}.  
This temperature dependence has shown a correction of several percent in the total pressure for some low Z systems 
in the warm dense matter regime \cite{sjostrom14b}.
In figure \ref{lu_xc} the effect on $Z^*$ of using a temperature dependent $F^{xc}$ is
plotted.  We have used the functional of Groth {\it et al} \cite{groth17}.  The top panel
shows the absolute change in $Z^*$ (in number of electrons per atom).  The effect is
generally quite modest, with $|\Delta Z^*| \lessapprox 0.1$. In panel (b) the relative change
in $Z^*$ is plotted. For $k_B > 50 $ eV the effect is $\lessapprox 1$ \%.  
At high temperatures
exchange and correlation effects become relatively small, compared to the kinetic energy, as the system becomes more ionized and
therefore more like an ideal non-interacting quantum electron gas.
At lower temperatures the relative effect of $F^{xc}$ is quite
large, approaching 50 \% at 0.1 eV.  However, in this region the absolute size of $Z^*$ is very small (see 
panels (c) and (d)).  The most significant effect is at $\approx 1$ eV and near solid density where
both the relative and absolute change in $Z^*$ are appreciable.  This is sometimes called the warm dense matter regime,
and is characterized by significant changes in electronic structure brought about by pressure ionization.

Finally, in figure \ref{lu_time} the wall-time taken to generate the data for figures \ref{lu_entropy} to \ref{lu_xc} is shown.
These wall-times are for the relativistic version.  The non-relativistic version is roughly a factor of 2 faster.
The choice of LDA exchange and correlation potential does not significantly affect the wall-time.  We ran {\texttt Tartarus} 
on an Intel\textregistered  Xeon\textregistered  CPU E5-2695 v4 (2.10GHz) with 18 physical cores.  18 instances of the serial code were run
and the same time, until all temperature/density cases were exhausted, for a total of 54 density points and 9 temperatures 
(486 temperature/density points total).  From the figure we see a dependence on temperature and density. The longest wall-times
occur where the plasma is very weakly degenerate and there is significant ionization (low density, high temperature).  
This is the regime where the Fermi-Dirac occupation factor has a slowly decaying energy tail, and where electrons in
high energy, orbial angular monentum states are weakly, but significantly, affected by the highly charged ion. 
Algorithmically, the wall-time increases here because
$l_{con}$ (section \ref{sec_ne}) is largest in this regime.  The majority of temperature/density points take less than five minutes.  It is worth noting that
any particular point could be made (possibly much) faster by tailoring the algorithm or number of grid or energy points.  However, 
these results were produced without any human interference; the same algorithm and numerical parameters were used everywhere.

We now turn to a comparison with a less approximate method.  We have used the plane wave DFT code {\texttt Abinit} \cite{abinit1, abinit2}
for calculations on hcp lutetium.  In contrast to the atomic sphere boundary
conditions and spherical model potential in \texttt{Tartarus}, periodic boundary 
conditions and a realistic 3d potential are used in {\texttt Abinit}, which is expected
to give more accurate results for the cold energy curve and the low energy
electronic spectrum. Our {\texttt Abinit} calculations use the projector augmented 
wave (PAW) method. The Lu PAW atomic data were generated using the Atompaw
code \cite{atompaw}. Our starting point for input parameters is the JTH v1.0
data set \cite{jollet14}. We reduced the PAW sphere radius from 2.5 to 
2.0 $a_B$ to avoid overlap at the highest compressions considered here.
All other radii, such as the pseudo-orbital matching radii and compensation
charge shape radius were scaled by 0.8. Atomic states up to $4d$ are treated 
as part of the frozen core. The LDA exchange correlation functional has been used for comparability 
with \texttt{Tartarus}. The plane wave cutoff in {\texttt Abinit} was
30 $E_h$. Cold energy calculations used ``cold smearing'' (occopt =4)
with smearing parameter 0.01 $E_h$ and an $8^3$ k-point mesh.
High temperature calculations used Fermi-Dirac occupation at the stated 
temperature with a $4^3$ k-point mesh. In the high temperature calculations,
the number of bands was set to 540, which results in occupation of the highest
band of $\sim 5 \times 10^{-4}$ at the highest temperature considered,
$T= 35$~eV.

\begin{figure}
\begin{center}
\includegraphics[scale=0.40]{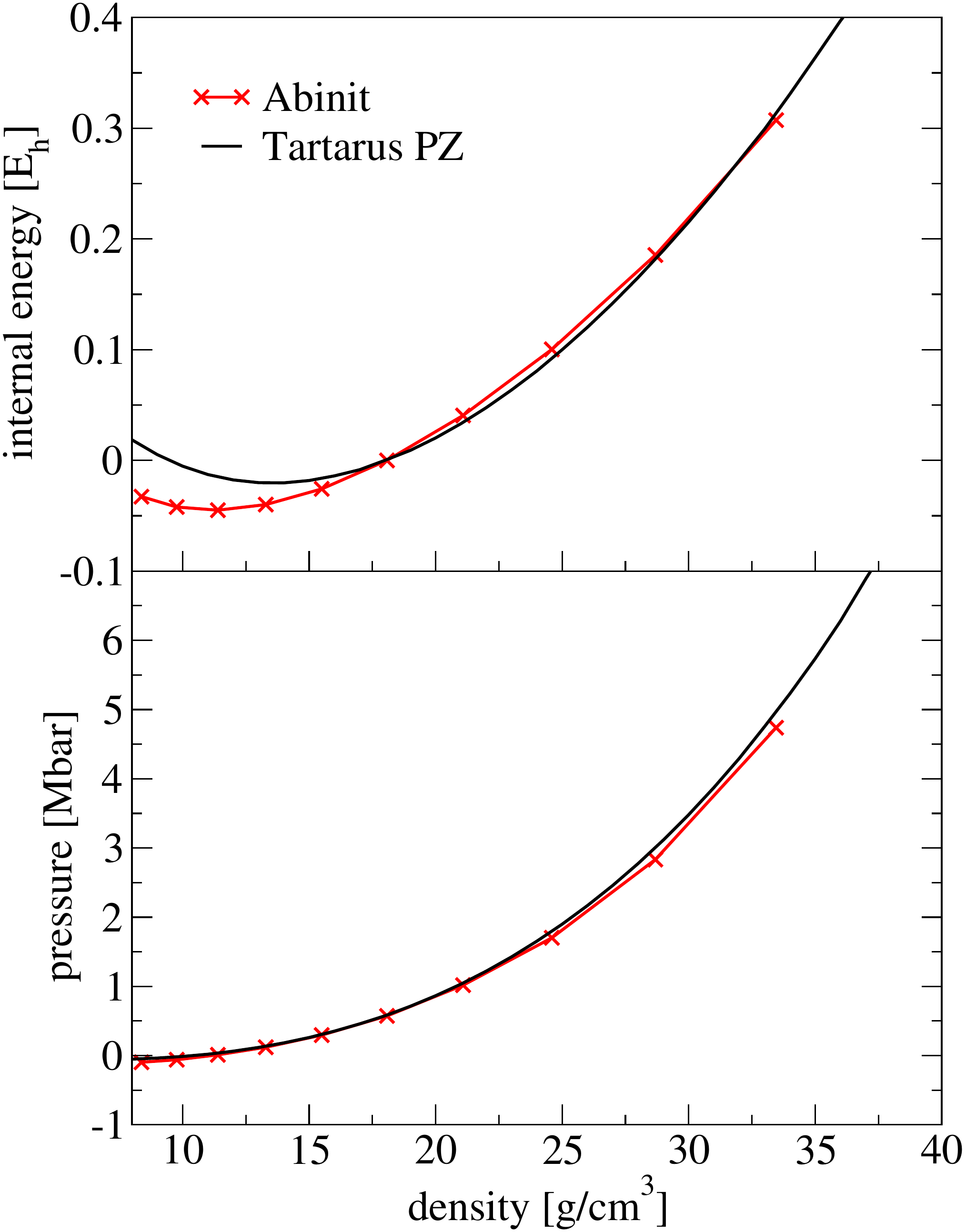}
\end{center}
\caption{(Color online) 
  Isotherm (0.0285 eV) for lutetium compared to the plane wave code {\texttt Abinit} using the HCP crystal structure.  
}
\label{lu_cold_curve}
\end{figure}

In constructing wide-range equations of state, an average atom model is often
used for the contribution of thermally excited electrons, with the cold energy 
and pressure subtracted. This allows for an empirical cold curve, or one
calculated with a more detailed electronic structure method, to be 
substituted.
In figure \ref{lu_isocore} we compare the thermal equation of state 
from {\texttt Tartarus} to {\texttt Abinit} for lutetium at solid density
for temperature from $\approx$ 0.25 eV to $\approx$ 25 eV.  Overall, there is a remarkable level of agreement
between the two methods.  For internal energy (panel (a)) some differences appear at the two highest temperature
points.  For these points we have noted the value of the Fermi-Dirac occupation factor as calculated in
{\texttt Tartarus} for the 4d state.  Clearly this state is beginning to be temperature ionized, indicating
that the frozen core approximation used in the {\texttt Abinit} calculations is near the limit of its validity, and
is likely the cause of the difference seen.  For entropy (panel (b)) small differences between the models are
apparent.  It is not surprising that the spherically symmetric average atom model that does not explicitly
account for crystal structure fails to exactly reproduce the less approximate plane wave code.  Nevertheless,
despite these approximations the level of agreement seen is very good.
Even at low temperature, where details of the low energy spectrum are important,
the two methods differ by well under $1 k_B$. The structure in the entropy
curves at $\sim 2$~eV is well reproduced by {\texttt Tartarus}.
For pressure (panel (c)) the agreement is again excellent, with the only significant differences appearing at high 
temperature, again likely due to the frozen core approximation in {\texttt Abinit}.

In figure \ref{lu_cold_curve} we compare {\texttt Tartarus} to {\texttt Abinit} for the cold pressure and energy.
The cold EOS is sensitive to details of chemical bonding, and we expect it to be the most challenging for the average atom model. 
The {\texttt Tartarus} calculation uses an electron temperature of 0.0285 eV, 
while {\texttt Abinit} uses Fermi surface smearing as described above. 
In the figure we can see that there are significant differences between
the models for internal energy, approaching 0.04 E$_h$ at normal density.  Perhaps more importantly is that the trends
as a function of density are not well reproduced by the simpler model.
It is worth noting that such absolute differences would not be apparent if plotted on the same scale as figure 
\ref{lu_isocore}.   The point being that while {\texttt tartarus} clearly gets large scale trends correct, smaller
scale trends may be incorrect.
\begin{figure}
\begin{center}
\includegraphics[scale=0.40]{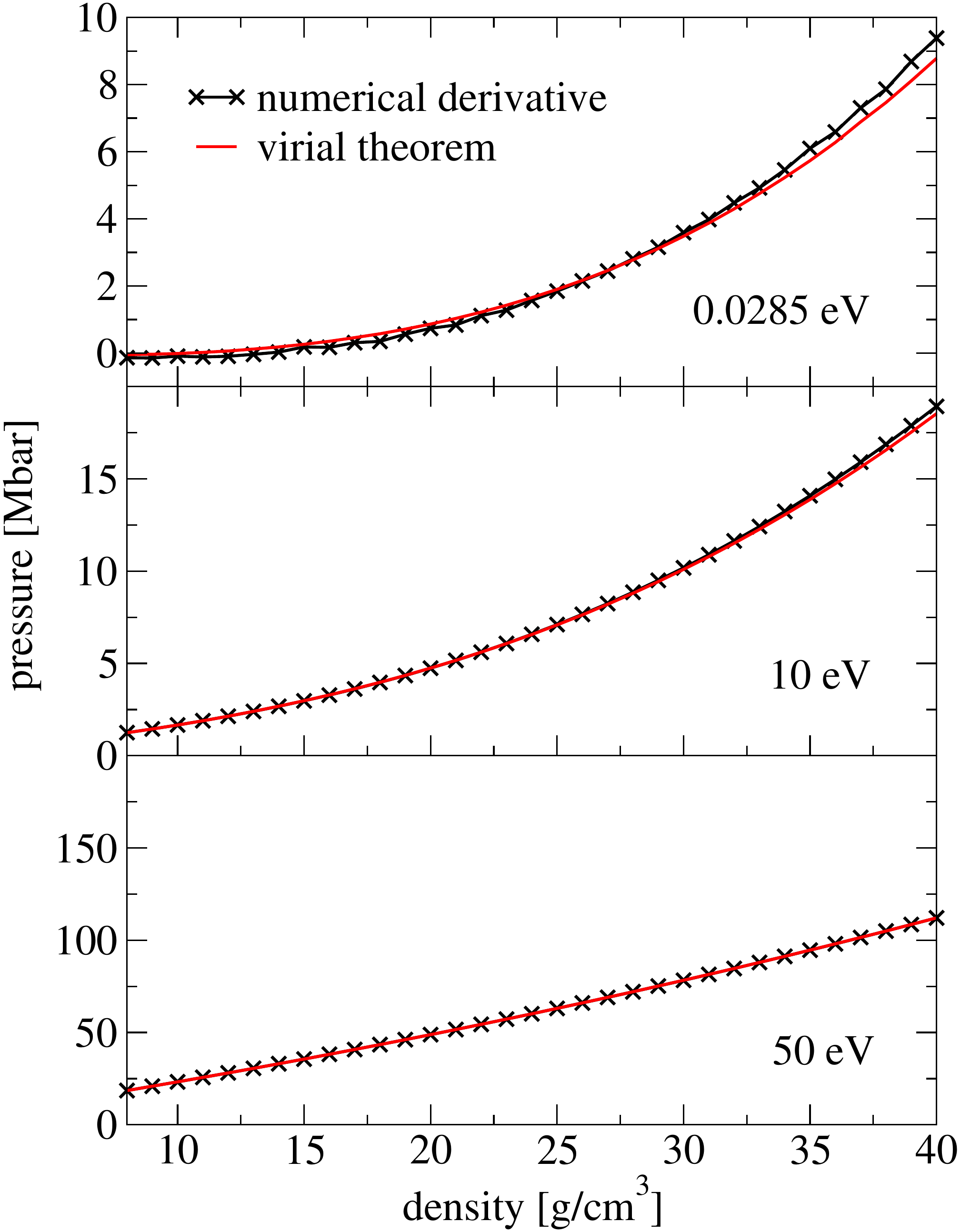}
\end{center}
\caption{(Color online) 
  Comparison of pressures from {\texttt Tartarus} for lutetium calculated with the Virial expression
  (\ref{pvir}) and via a numerical differentiation of the free energy $P=-\partial F/\partial V|_T$.
} 
\label{lu_thermocon}
\end{figure}

For pressure, figure \ref{lu_cold_curve}, the agreement is reasonable on the scale of the figure.  
The pressure shown is calculated using the Virial expression, equation (\ref{pvir}).  It is also
possible to calculate the pressure by taking a numerical derivative of the free energy $F$
\begin{equation}
P = - \left. \frac{\partial F}{\partial V}\right|_T
\end{equation}
As is well documented \cite{gill17, piron11}, the physical model that {\texttt Tartarus} uses
does not guarantee the these two pressures will be identical.  In figure \ref{lu_thermocon}
we show the pressure calculated both these ways for three isotherms of lutetium.  For the cold curve 
(0.0285 eV) significant differences are observed.  By 10 eV the differences are largely gone
but show up at the highest densities.  At 50 eV the agreement between the two pressures is very
good.  Generally differences appear where oscillations in the electron density have not
died out by the sphere boundary.  Such oscillations are a consequence of a sharp
Fermi-Dirac distribution which occurs in degenerate systems and are
called Friedel oscillations.  The figure reflects this: the larges differences are seen for
the most degenerate systems (i.e. low temperature and high density).

Such an inherent thermodynamic inconsistency may or may not be problematic depending on
the application of the model.  A practical solution is to just use the free energy
to generate the entire EOS through numerical derivatives.  Such an approach generates other
problems, principally that the free energy must be smooth enough for the derivatives
to be accurate.  For many applications however, such inconsistency is not
particularly problematic.  We note that a thermodynamical consistent average atom
is possible \cite{blenski2, piron11}.

\section{Conclusions}
We have presented a detailed discussion of the physics model and numerical implementation
of the {\texttt Tartarus} average atom code.  The model is based on a hybrid orbital and Green's
function implementation and the advantages of such a scheme are presented.  A numerically
efficient method of solving the self consistent field problem is also given.  It is hoped
that this presentation may guide others in their own implementations.

We then focus on the application of the model to a lutetium plasma for a wide range of conditions.
We use this example to explain concepts such as broadening of the density of states in
the complex energy plane, and prediction of ionization.  The effect of relativity on the wide ranging
EOS is also presented.  It is found that relativity is generally a small effect, but is important for
internal energy, and for entropy at low temperature and density.

The effect of finite temperature exchange and correlation potentials is also investigated.  It is found
the effect is generally small, but becomes relatively significant for warm dense matter conditions,
i.e. near normal density and temperature around 1 eV.

A comparison of the model to a more physically realistic model at low temperatures reveals the {\texttt Tartarus}
model is generally in very good agreement with the more physically accurate model, but that smaller scale
deviations are apparent.

Some oddities of the model are discussed.  We find that an increase in entropy near normal density 
along at low temperature isotherm corresponds to a region where pressure decreases as temperature
increases, for fixed density.  This artifact of the model occurs in a small region, at low temperature
where the material is transitioning from metal to non-metal.  Also, thermodynamic inconsistency
for highly degenerate materials is discussed.

In summary, the hybrid orbital/Green's function approach to the average atom model was found to be
very stable numerically and is recommended for future implementations.


\section*{Acknowledgments}
We are grateful to B. Wilson for beginning our interest in the Green's function approach.
This work was performed under the auspices of the United States Department of Energy under contract DE-AC52-06NA25396
and LDRD number 20150656ECR.

\bibliographystyle{unsrt}
\bibliography{phys_bib}

\end{document}